%% file: main.tex
\documentclass[review]{elsarticle}

\usepackage{lineno,hyperref}
\modulolinenumbers[5]

\usepackage[table,xcdraw]{xcolor}
\usepackage{times}
\usepackage{enumerate}
\usepackage{graphicx,adjustbox}
\usepackage{hyperref}
\usepackage{amsmath,amssymb}
\usepackage{blindtext}
\usepackage{multicol}
\usepackage{makecell}
\usepackage{graphicx,color}
\usepackage{multirow}
\usepackage{color}
\usepackage{rotating}

\usepackage{diagbox}
\usepackage{array}

\usepackage{booktabs} 
\usepackage{amsmath}
\usepackage{wrapfig}
\usepackage{subcaption}
\usepackage{epsfig}
\usepackage{epstopdf}
\usepackage{float}
\usepackage{graphicx}
\usepackage{chngcntr}
\usepackage[T1]{fontenc}
\usepackage{lmodern} 
\usepackage{array}
\usepackage[abs]{overpic}
\graphicspath{{./figure/}}

\usepackage{amscd}
\usepackage{graphicx}
\usepackage[all]{xy}

\journal{Journal of \LaTeX\ Templates}

\begin{document}

\begin{frontmatter}

\title{Source Printer Identification using Printer Specific Pooling of Letter Descriptors}
\author{Sharad Joshi\fnref{myfootnote1}}
\author{Yogesh Kumar Gupta\fnref{myfootnote2}}
\author{Nitin Khanna\fnref{myfootnote3}}
\fntext[myfootnote1]{Electrical Engineering, Indian Institute of Technology Gandhinagar\\
Email: sharad.joshi@alumni.iitgn.ac.in}
\fntext[myfootnote2]{Mathematics, Indian Institute of Technology Gandhinagar\\
Email: yogesh.gupta@alumni.iitgn.ac.in}
\fntext[myfootnote3]{Electrical Engineering and Computer Science, Indian Institute of Technology Bhilai (IITBH), Multimedia Analysis and Security (MANAS) Lab, Room no. B010, GEC Campus, Sejbahar, Raipur, Chhattisgarh, India - 492015\\
Email: nitin@iitbhilai.ac.in}

\input{Sections/S0_Abstract.tex}


\end{frontmatter}


\input{Sections/S1_Introduction.tex}
\input{Sections/S21_Related.tex}
\input{Sections/S22_PSLTD.tex}
\input{Sections/S3_Proposed.tex}
\input{Sections/S4_Expererimental.tex}
\input{Sections/S5_Conclusion.tex}

\section*{Acknowledgment}
This material is based upon work partially supported by the [http://www.serb.gov.in]Department of Science and Technology (DST), Government of India under the Award Number ECR/2015/000583 and supported by Visvesvaraya PhD Scheme, MeitY, Govt. of India MEITY-PHD-951.
Any opinions, findings, and conclusions or recommendations expressed in this material are those of the author(s) and do not necessarily reflect the views of the funding agencies.


\bibliography{main}

\end{document}

%% file: Sections/S0_Abstract.tex
\begin{abstract}
The digital revolution has replaced the use of printed documents with their digital counterparts.
However, many applications require the use of both due to several factors, including challenges of digital security, installation costs,  ease of use, and lack of digital expertise.
Technological developments in the digital domain have also resulted in the easy availability of high-quality scanners, printers, and image editing software at lower prices.
Miscreants leverage such technology to develop forged documents that may go undetected in vast volumes of printed documents.
These developments mandate the research on creating fast and accurate digital systems for source printer identification of printed documents.
We extensively analyze and propose a printer-specific pooling that improves the performance of printer-specific local texture descriptor on two datasets.
The proposed pooling performs well using a simple correlation-based prediction instead of a complex machine learning-based classifier
achieving improved performance under cross-font scenarios.
The proposed system achieves an average classification accuracy of 93.5\%, 94.3\%, and 60.3\% on documents printed in Arial, Times New Roman, and Comic Sans font types respectively, when documents printed in only Cambria font are available for training.
\end{abstract}

%% file: Sections/S1_Introduction.tex
\section{Introduction}
\label{sec1}
Printed documents have been used traditionally for record-keeping, certification, and communication purposes.
The digital revolution has propelled the use of digital documents in place of printed documents.
However, many critical applications use printed documents like financial dealings, judicial process, administrative record keeping and communication, and certifications.
The frameworks for many such applications require a combination of printed documents, scanned versions, and digital documents.
This co-existence of printed and digital documents results from multiple reasons, including ease of use, operating cost, lack of digital expertise, and open digital security challenges.
Technological developments have allowed the easy availability of good quality printers, scanners, and image-editing software.
These are routinely misused by potential miscreants to create forged documents.
The vast volume of printed documents and the evolving forging technology makes it challenging to analyze and detect forged documents and warrants the use of fast and accurate digital systems to analyze the authenticity of printed documents.
The source printer's information can provide important clues for forensic analysis of printed documents~\cite{chiang2009printer, ferreira2017data,joshi2018single,joshi2019source}.

Traditional methods use chemical techniques to investigate the chemical composition of toner spread on a paper. Chemical processes use spectroscopy~\cite{braz2013raman} and 
x-ray~\cite{chu2013forensic} to discover a connection in seized documents.
These methods require laboratory equipment and an expert to examine the samples. 
Also, these techniques require a significant amount of time and can damage the printed document in question.
In sharp contrast, digital methods convert the printed documents into their digital counterpart using a reference scanner.
Source printer identification with digital techniques has been gaining much attention.
All the analysis is carried out in the digital domain; thus, it is faster and automatic.
Two major digital approaches for source printer identification are discussed in the literature, namely, extrinsic (active) and intrinsic (passive)~\cite{chiang2009printer}.
An extrinsic signature approach is an active approach in which a user embeds an extrinsic signal in the printed document before or during the printing process. 
However, they require active access to the printer, and a sophisticated signal embedding mechanism needs to be integrated with the printer~\cite{Chiang2011}.
Such methods are complicated and costly for the large volume of text documents printed by general-purpose consumer-grade printers.
On the contrary, passive approaches rely on visually imperceptible printer artifacts induced intrinsically during the printing process~\cite{mikkilineni2005printer}.

In this paper, we work on source printer identification for text documents using an intrinsically induced printer signature.
The existing methods process all the printed letters uniformly during the computation of the printer signature model for source printer identification in a close-set scenario (i.e., the user needs to know the set of all possible printers in advance).
Our proposed method is based on the hypothesis that the printed letters exhibit location-specific variations due to the electrophotographic printing process's characteristics~\cite{ali2003intrinsic}.
We introduced the printer-specific local texture descriptor (PSLTD) in~\cite{joshi2019source}, a state-of-the-art handcrafted method for identifying the source printer using scanned images of printed documents.
Most existing methods (including deep learning-based~\cite{ferreira2017data,joshiIcassp2018}) do not perform well when the font type of letters in printed documents under test is not available in training.
We term this as the cross-font scenario~\cite{joshi2019source}.
PSLTD outperforms all existing methods under a cross-font scenario.

We analyze the variation of PSLTD-based printer signature concerning the location of printed letters across the document.
We show that the distribution of printer artifacts correlates strongly with printed letters' location on the document (Figures~\ref{fig:corr_col}).
We introduce a printer-specific pooling technique that allows the prediction of source printer.
We further improve the performance of PSLTD under a cross-font scenario using the proposed pooling technique.
We show the efficacy of the proposed method using a correlation technique (Figure~\ref{fig:overall_pipeline_corr}).
The significant highlights of this work are as follows:
\begin{itemize}
    \item We provide an extensive analysis of printer signature variations across a printed document.
    \item We introduce a location-based pooling technique that improves the performance of PSLTD for source printer identification.
    \item We show the efficacy of the proposed pooling technique using a correlation-based prediction method instead of a complex classifier, thus paving classifier-independent prediction.
    \item The proposed method performs better than state-of-the-art methods under the cross-font scenario achieving an average classification accuracy of 93.5\%, 94.3\%, and 60.3\% on documents printed in Arial, Times New Roman, and Comic Sans font types when documents printed in only Cambria font are available for training.
\end{itemize}
The paper consists of the following sections.
Section~\ref{sec:Related} briefly describes the related literature of intrinsic signature-based techniques for classifying the source printer of printed text documents.
The details of our proposed pooling have been specified in Section~\ref{sec:Proposed}. 
An extensive set of experiments are used to examine the effectiveness of the proposed method.
The description and results of the proposed approach have been discussed in Section~\ref{sec:Experimental}. 
Finally, we draw out conclusions from this work and the directions for future work in Section~\ref{sec:Conclusion}.

%% file: Sections/S21_Related.tex
\section{Related Works}
\label{sec:Related}
The problem of Source printer classification has gained much attention in the past decade~\cite{chiang2009printer}.
This work is based on an intrinsic signature, so we focus only on them.
Our method also extracts printer signature from images of printed letters, so we only discuss methods related to text documents.
One of the early reliable printer signatures was based on the banding phenomenon, i.e., the appearance of light and dark lines perpendicular to the direction of the paper movement inside the printer~\cite{chiang2009printer}.
It is a promising approach but requires documents images scanned at a very high resolution (2400 dpi).

From banding, the researchers moved towards texture-based methods.
These methods are suitable for documents scanned at lower resolutions (300-600 dpi).
The texture patterns created by the distribution of intensity values observed in the scanned document image form the basis for these methods.
These variations are visually imperceptible to human eyes at commonly used font sizes.
Nevertheless, their zoomed versions depict texture variations.
These methods follow the traditional pattern recognition pipeline in which features are extracted from letter images followed by learning a suitable classifier model.
The learned model is capable of predicting the source printer labels for each letter image.
The majority voting over all the predicted letter labels provides the prediction for the whole document under test.
The early methods in this category were developed for a specific letter type like `e', which is the most frequently occurring letter in the English language.
The features include gray-level co-occurrence matrices (GLCM)~\cite{Mikkilineni2005}, GLCM extended in multiple directions, and scales~\cite{ferreira2015laser}, convolutional texture gradient filter (CTGF) based on filtering textures with a specific gradient~\cite{ferreira2015laser}, a combination of the discrete wavelet transform (DWT) and GLCM based features~\cite{Tsai2014}, a combination of features obtained after applying a spatial filter, Wiener filter, and Gabor filter~\cite{tsai2015japanese}, and GLTrP-based features~\cite{joshi2018single}.
A recent method proposes a decision-fusion model-based approach for source printer classification~\cite{tsai2018decision}.
All these methods extract features that learn a model for a specific letter type except~\cite{joshi2018single}, which introduces a single-classifier approach that learns a single model for all letter types printed on a document.
The use of all letters increases training samples, thus lowering the requirement on the number of training documents and also increasing predicted labels for a document under test
to achieve good performance. 

Elkasrawi and Shafait~\cite{Elkasrawi2014} and the authors in~\cite{joshiIcassp2018} proposed noise residual-based features.
Kee and Farid~\cite{kee2008printer} followed a different strategy and proposed a method based on the estimation of a printer profile.
They create the profile using a mean character image (of letter type `e'), and top \textit{p} eigenvectors (obtained by PCA) is generated for each printer.
The mean character image is obtained from all occurrences of letter `e' extracted by convolving each character image with a user-selected reference character image.
Zhou \textit{et al.}~\cite{zhou2015text} proposed an interesting 
text-independent approach to identify source printer based on a piece of patented equipment to scan fine textures on printed pages.
More recent work proposed a method for source printer identification using document images acquired using a reference smartphone camera~\cite{joshi2020source}.

Two methods have also been proposed using convolutional neural networks (CNN)~\cite{ferreira2017data,joshiIcassp2018}.
They replace the task of handcrafted feature extraction with a model learned by letter images.
Both methods work using specific letter types (`e' and `a').
None of the techniques discussed so far have been evaluated for cross-font scenario (i.e., font type of documents in testing is not available during training.)
This is crucial as there are so many font types available, and it is increasingly difficult to expect that all font types are available during training.
A recent method based on printer-specific local texture descriptor (PSLTD)~\cite{joshi2019source} tries to address this challenge.
A completely different category is characterized by geometric distortion-based approaches~\cite{bulan2009geometric, wu2009printer}, which rely on features from translational and rotational distortions of printed text relative to its reference soft copy~\cite{hao2015printer, shang2015, jain2017passive}.
A detailed literature review has been covered in~\cite{ferreira2015laser,ferreira2017data,joshi2018single}.

%% file: Sections/S22_PSLTD.tex
\section{Printer Specific Local Texture Descriptor}
\label{sec:PSLTD}
Printer specific local texture descriptor~\cite{joshi2019source} is one of the most state-of-the-art digital systems for source identification of printed documents.
The significant advantage of the PSLTD-based method is its performance under a cross-font scenario in which it outperforms existing methods by huge margins.
PSLTD is a novel handcrafted feature extracted from each letter image capable of learning a single discriminative classifier model for all letter types.
It belongs to the family of local binary patterns~\cite{Ojala2002}.
The design of our PSLTD emphasizes on preprocessing,
thresholding and quantization, and encoding and regrouping
stages of local binary feature extraction.
It introduces a novel encoding and regrouping strategy built upon small linear structures of 3 $\times$ 1 shaped structures.
The main hypothesis of PSLTD is that the distribution of such linear structures in small overlapping 3 $\times$ 3 patches can effectively encode and group local binary patterns.
The method observes linear-shaped groups of pixels that exist around the center pixel at all possible orientations, i.e., horizontal (0$^{\circ}$), vertical (90$^{\circ}$), forward slant (45$^{\circ}$), and backward slant (135$^{\circ}$).
Each 3 $\times$ 3 patch is assigned one or more line orientations based on intensity and gradient direction similarity.
A pent-pattern vector is calculated from contiguous 3 $\times$ 3 sized patches, which is converted into five binary pattern vectors (BPVs).
These BPVs from a local patch are mapped to multiple normalized pattern occurrence histograms based on uniformity~\cite{Ojala2002} by encoding and regrouping them by their line orientations.
The detailed mathematical description of PSLTD is beyond the scope of this paper. For more details, please refer~\cite{joshi2019source}.
The final PSLTD is a feature vector of 10502 dimensions comprising of four major components.
The first group comprises normalized histograms obtained using intensity and gradient direction similarities as encoding and regrouping strategies. It is denoted by $\overrightarrow{F1}$ and is of 4425 dimensions.
Similarly, the second and third groups are formed by considering only intensity similarity and only gradient direction similarity to encode and regroup the BPVs.
They are denoted by $\overrightarrow{F2}$ and $\overrightarrow{F3}$ and are of 1475 and 4425 dimensions, respectively.
The last component is formed by binary magnitude pattern vector (BMPV) denoted by $\overrightarrow{F}_{BMPV}$ and of 177 dimensions.
\begin{figure*}[t!]
    \centering
    \includegraphics[width = \textwidth]{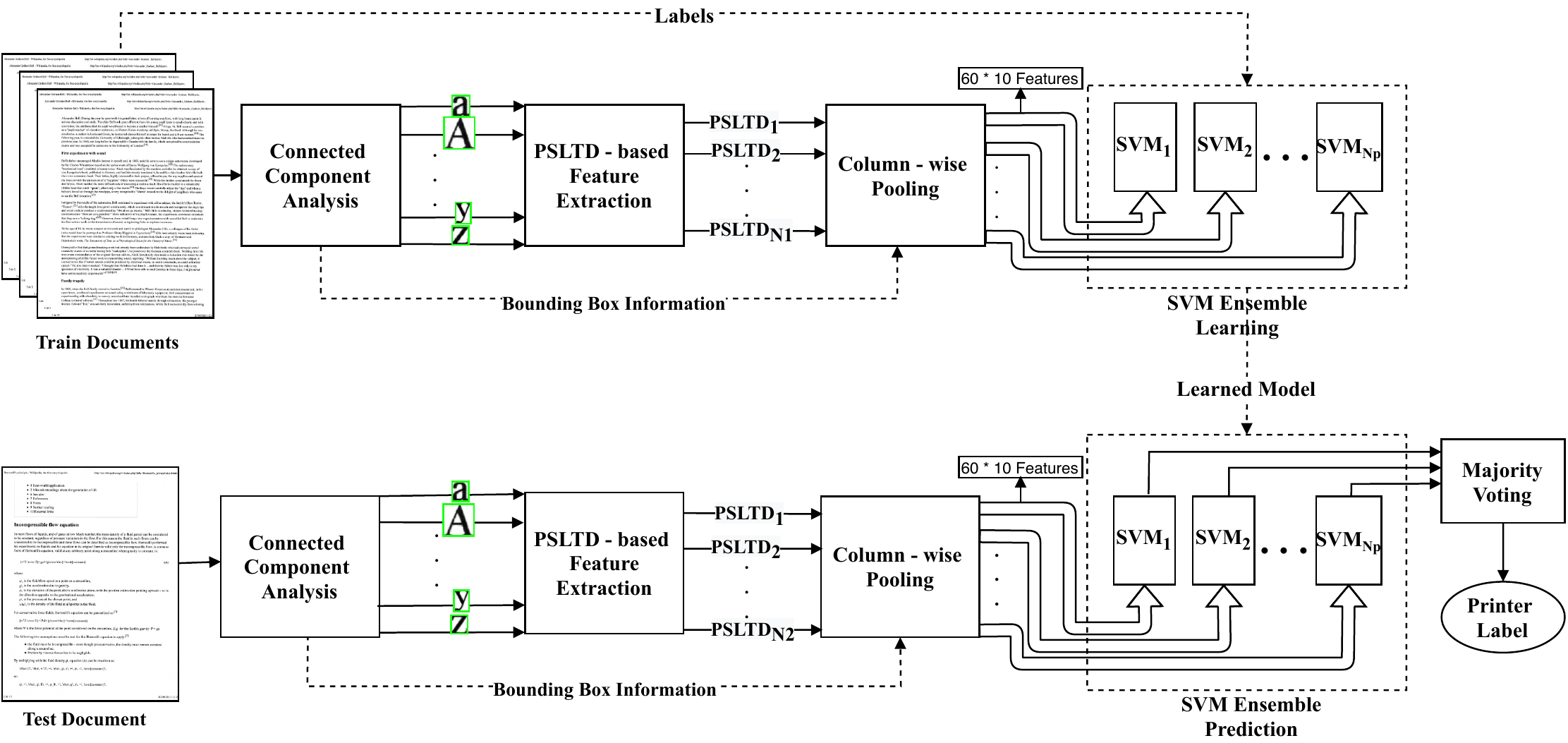}
    \caption{ Overall pipeline of the proposed approach using classifier-based prediction (for documents scanned using 8-bit depth intensity values).}
    \label{fig:overall_pipeline_svm}
\end{figure*}
\begin{figure*}[t!]
    \centering
    \includegraphics[width = \textwidth]{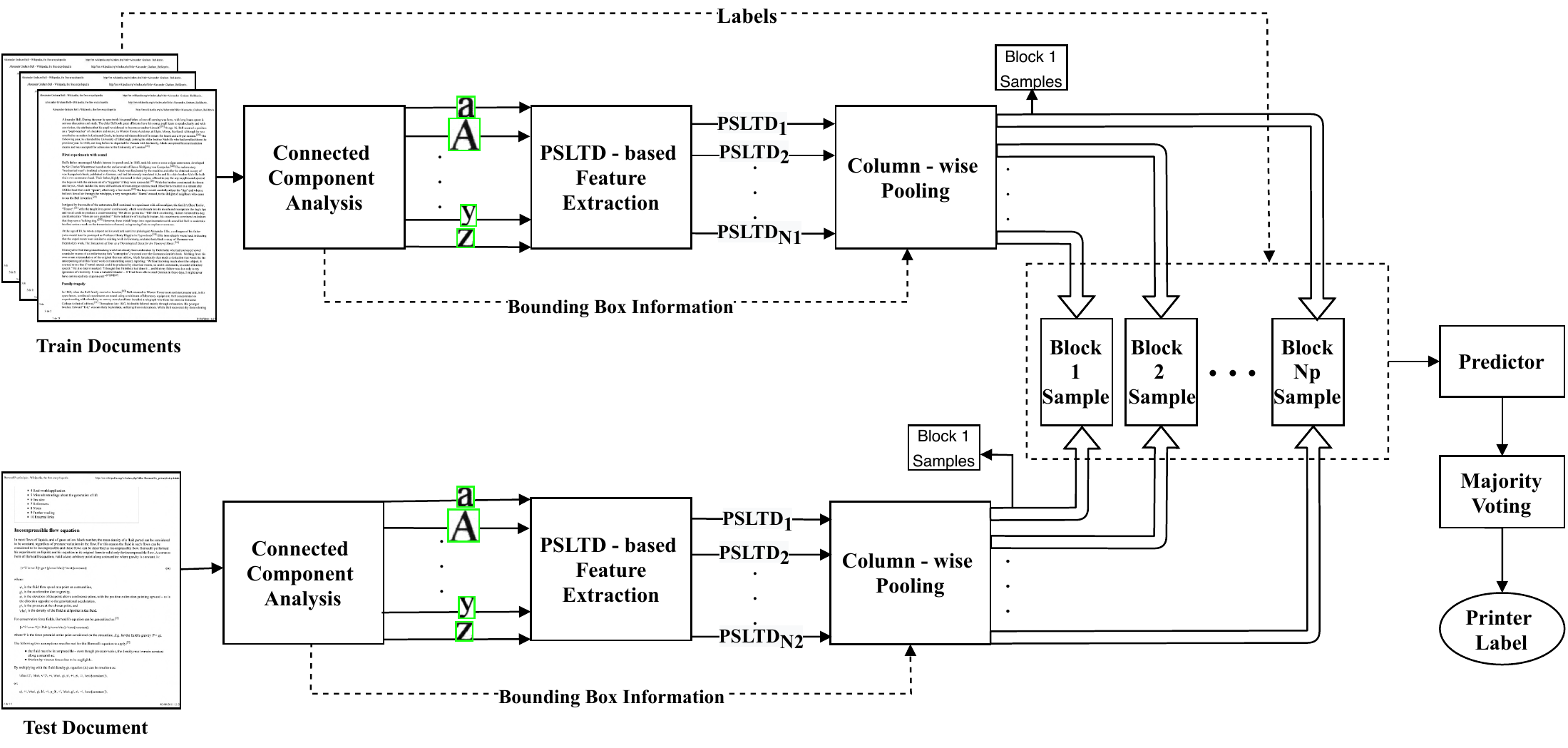}
    \caption{ Overall pipeline of the proposed approach using classifier-based prediction (for documents scanned using 16-bit depth intensity values).}
    \label{fig:overall_pipeline_corr}
\end{figure*}

%% file: Sections/S3_Proposed.tex
\section{Proposed Method}
\label{sec:Proposed}
In this work, we analyze the change in PSLTD-based printer signature~\cite{joshi2019source} concerning the location of printed letters across the document.
Based on the analysis, we introduce a printer-specific pooling technique that enhances source printer classification accuracy.
We try two types of pooling-topology, i.e., columns and grids.
Our proposed system's input is document images of hard copy printed documents acquired using a reference scanner.
The choice of reference scanner does not impact the system performance as long as it is kept constant for all the documents.
The proposed system comprises of four significant steps: 1) connected component extraction, 2) PSLTD-based feature extraction, 3) and printer-specific pooling, and 4) classification.
The general pipeline remains the same as in~\cite{joshi2019source}. 
The novel step is our proposed pooling strategy, which enhances classification performance.
Figure~\ref{fig:overall_pipeline_svm} depicts an overall pipeline of the proposed system using classifier-based prediction.
For images scanned and saved using 16-bit intensity values (our dataset~\cite{joshi2018single}), the proposed method also allows prediction of source printer using a correlation technique instead of a complex classifier-based prediction.
as shown in Figure~\ref{fig:overall_pipeline_corr}.

\subsection{Connected Component Extraction}
\label{subsec:conn_comps}
The first step is the extraction of individual letter images using connected component analysis from a printed document~\cite{joshi2018single}.
We binarize the document image followed by bounding box detection around each connected component following the procedure in~\cite{joshi2018single}.
Our interest lies in letter images. However, the system also finds some spurious components due to punctuation and noise.
We selectively filter and remove bounding boxes that are too small or large in area and dimension.
Such filtering parameters are empirically selected, depending on the text's characteristics printed in the document.
Specifically, for our dataset~\cite{joshi2018single}, we remove all components larger than four times the median of areas of all components or smaller than 0.5 times the median of areas of all components on a document.
For the publicly available dataset~\cite{ferreira2015laser}, we also need to remove components with a width smaller than 15 and larger than 90 pixels and components of height smaller than 30 and larger than 100 pixels.
This technique also removes the dots occurring in letter types `i' and `j'.
Our method performs this step on all train and test documents.

\subsection{Feature Extraction}
\label{subsec:Feat_ext}
The feature extractor stage input is the letter images extracted using the bounding boxes obtained by connected components analysis.
A PSLTD is extracted from each connected component (C), i.e., letter images of train and test document.
However, instead of using the full length PSLTD, we show that its approximated smaller length version is sufficient.
In particular, we use the $\overrightarrow{F1}$ and $\overrightarrow{F}_{BMPV}$ components.
This approximation reduces the feature vector dimension to 4602 dimensions, reducing complexity and time to run so many experiments.

\subsection{Printer Specific Pooling}
The feature pooling step is usually inserted in modern visual classification techniques.
Pooling aggregates local features into a static value via some pooling operation~\cite{2011Fenggeometric}.
Our work's primary hypothesis is that the artifacts induced by a printer are not identical for the entire printed document.
It is a well-established observation that the occurrences of the
same letter on a document printed in one go by a single printer
may have different intensity distribution~\cite{joshiIcassp2018}.
One possible reason for this observation is the complex electro-mechanical parts and circuitry involved in the printing process.
The electrophotographic printing process follows a line by line approach from top to bottom~\cite{ali2003intrinsic}.
The various steps in the printing process occur in the same fashion.
So, we expect that the combined artifacts induced by all these steps will contain particular variation while moving perpendicular to the printer process direction (i.e., horizontal row-wise fashion for the document image of a printed document). In contrast, we expect that the variation along the process direction must be smaller.
However, some other sources may also introduce variations that may not occur identically in a horizontal row-wise fashion.
We analyze our hypothesis using two techniques, namely, column pooling and grid pooling.

\subsubsection{Column Pooling}
\begin{figure}
    \centering
    \includegraphics[width = \textwidth]{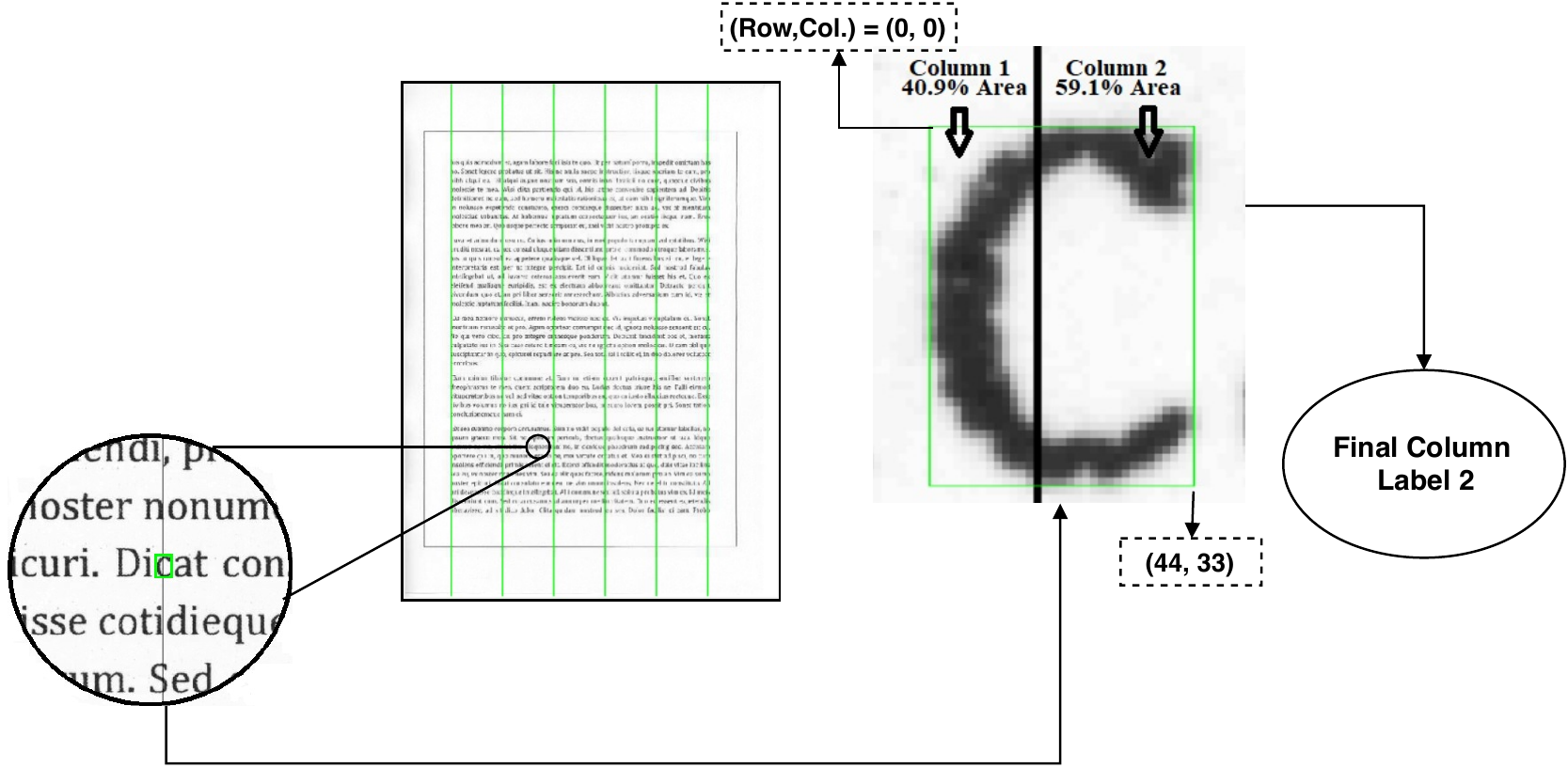}
    \caption{Illustration of column pooling and column number assignment for boundary letters on the basis of area.}
    \label{fig:col_select}
\end{figure}

\begin{figure*}[!ht]
    \centering
    \includegraphics[width = \textwidth]{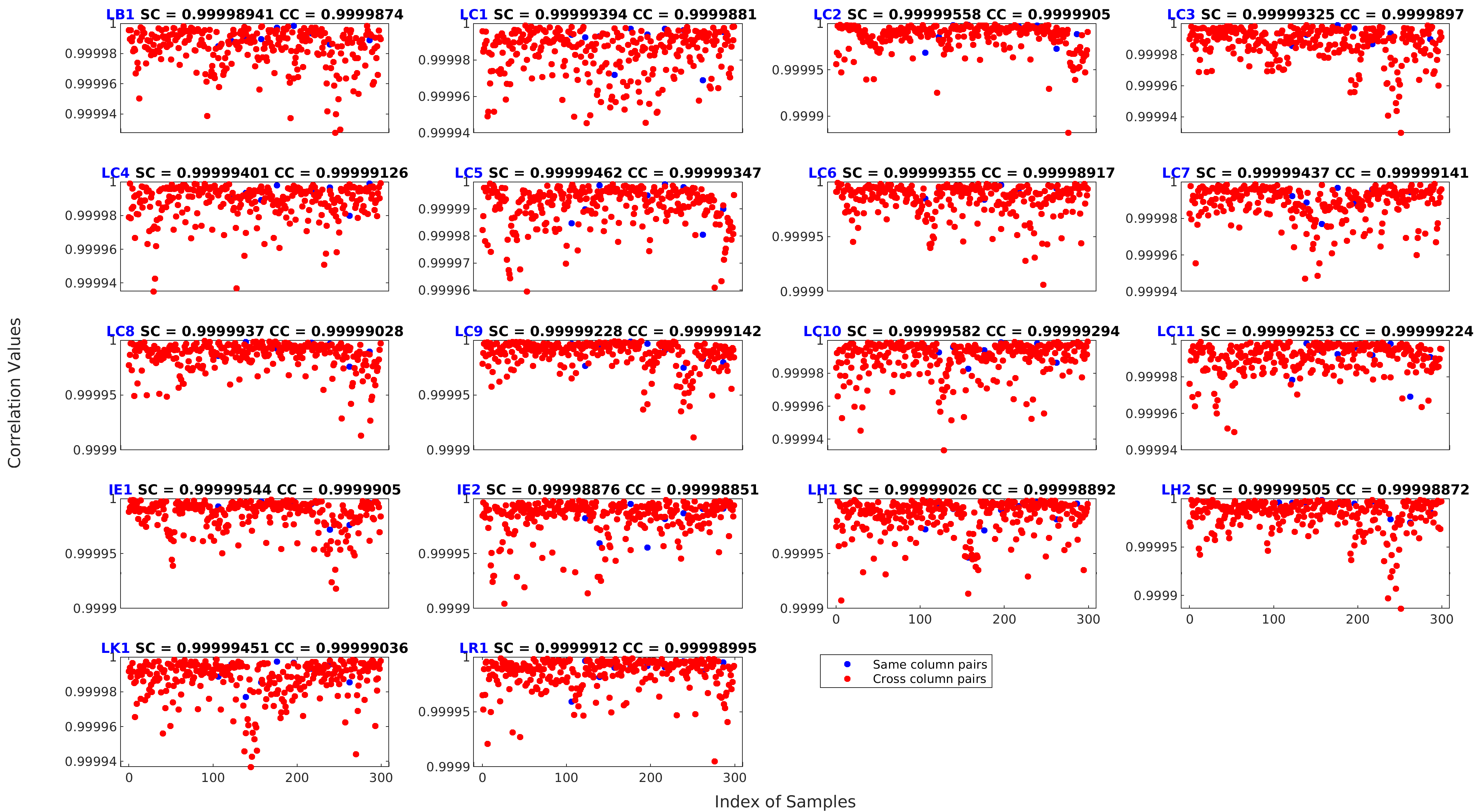}
    \caption{Correlation values of feature vectors obtained from letters in same and different columns for all printers in dataset DB2\cite{joshi2018single}. SC and CC denote the median values obtained by correlating same and cross-column PSLTD pairs, respectively.}
    \label{fig:corr_col}
\end{figure*}

In this pooling strategy, we divide each printed document into a specific number of vertical columns (i.e., along the printer process direction) using some fixed rules.
At first, we estimate the empty horizontal margin space on both the right and left sides of the printed text as it does not contain any printed letters.
We estimate the start and endpoints of printed text using the minimum and maximum value of horizontal coordinates, i.e., row-coordinates.
Specifically, for each document, we sort the bounding boxes of connected components based on their row-coordinates.
Then, we calculate the median values of 1\% of the smallest (i.e., left-most) and largest (i.e., right-most) coordinates of the bounding boxes of connected components.
These two median values establish the start and endpoints of printed text denoted by $P_{1}$ and $P_{2}$, respectively.
We chose the median operator to avoid the effect of any spurious components that may still pass the filtering process of the connected component extraction stage (Section~\ref{subsec:conn_comps}).
The first column's start is a 1-pixel wide boundary passing through $P_{1}$ and parallel to the document image's vertical boundary.
Similarly, the last column's end is a 1-pixel wide boundary passing through $P_{2}$ and parallel to the document image's vertical boundary.
The printed text area's start and endpoints allow the horizontal margin space to be removed as it does not contain any data of interest.
The boundaries of all columns are equispaced and parallel to the vertical boundary of the document image.
The spacing between them is termed as column-width (C$_{W}$), which is defined as the ratio of horizontal space covered by printed text to the number of columns (N$_{c}$) as follows.
\begin{equation}
    C_{W} = \frac{max\{c_{row}\} - min\{c_{row}\}}{N_{c}}
\end{equation}
Here, $c_{row}$ denote the set of row coordinates of all connected components. $max\{.\}$ and $min\{.\}$ denote the functions to compute maximum and minimum values.
Each column's width depends upon the size of the document and side margins (i.e., left and right margin).
For example, in our dataset~\cite{joshi2018single},
the average size of a scanned document image is 70192 $\times$ 5100 pixels with right and left margin of 887 and 638 pixels, respectively.
So,  for $N_{p}$ = 15, the width of each column would be 238 pixels.
We assign a column to each connected component in a document image using the following rules.
\begin{itemize}
    \item If a connected component (characterized by its bounding box) is printed entirely inside a column, it is assigned to that column.
    \item If a connected component occurs at the intersection of two columns, we compute the area covered by each column's component. The column containing the larger area of that component is assigned to that component, as described in Figure~\ref{fig:col_select}.
    \item If a connected component occurs at the intersection of two columns with an equal area in both columns, we assign the left-most column number.
\end{itemize}
We pool the PSLTD of a group of N$_{p}$ letters assigned to a column by average pooling resulting in N$_{p}$ feature vectors. Here, N$_{p}$ is termed as the pooling parameter.
The hypothesis behind post-extraction pooling (PoEP) is that it reduces the variance among PSLTDs extracted from the same page~\cite{joshi2019source}.
We expect PoEP to eliminate undesired noise as the pooled feature vectors represent all samples of a column.
We visualize the comparison of correlation values obtained by the same column and cross column pairs in Figure~\ref{fig:corr_col}.
The median correlation values of the same column (SC) pairs are consistently higher than those for cross column (CC) pairs across all printers.

\subsubsection{Grid Pooling}
We further analyze the variation of our PSLTD printer signature across the document by dividing the document into N$_{w}$ $\times$ N$_{h}$ grids.
We hypothesize that the printer signature may also vary within each column.
The general procedure of estimating the empty margin space and removing it remains the same as in column pooling.
In addition to removing horizontal space, we also need to remove the space at the top and bottom of the printed text area (vertical margin) to determine the top-most grids' horizontal boundary.
Each grid has a 1-pixel wide boundary determined using the estimated printed text area.
The dimensions of each grid (G$_{W}$ $\times$ G$_{H}$) are calculated from the horizontal and vertical space covered by the printed text area.
\begin{equation}
    G_{W} = \frac{max\{c_{row}\} - min\{c_{row}\}}{N_{w}}
\end{equation}
\begin{equation}
    G_{H} = \frac{max\{c_{col}\} - min\{c_{col}\}}{N_{h}}
\end{equation}
c$_{col}$ denote the set of column coordinates of all connected components.
For an 8 $\times$ 8 grid, each grid consists of approximately 665 $\times$ 446 pixels, and there are approximately 39 letters in one grid.
We assign a grid to each connected component (i.e., printed letter) similar to column pooling.
The PSLTDs of all letters in a grid are average pooled into a single feature vector.

\subsection{Prediction}
The prediction stage uses the pooled feature vectors to predict the source printer label corresponding to each feature vector.
We hypothesize a similarity in the printer signature for each block, i.e., either a column or a grid.
This hypothesis allows us to train and test letter samples separately for each block.
\subsubsection{Using SVM}
The PSLTDs of all letters in a block for all training documents are used to learn a single classifier model.
For column pooling, we train N$_{p}$ SVM classifier models, one for each column.
SVM is a standard classifier used by most existing methods for source printer identification~\cite{ferreira2015laser,joshi2018single,joshi2019source}.
During testing, the p$^{th}$ classifier model is used to predict the source printer labels of N$_{p}$ pooled feature vectors corresponding to a group of letters in the p$^{th}$ column (p $\in$ \{1,2,...,N$_{p}$\}) as shown in Figure~\ref{fig:overall_pipeline_svm}.
Finally, a majority vote on predicted labels of all groups of letters in the document provides the printer label for the printed document under test.
Grid pooling uses a similar strategy.
The major difference is that there is only a single pooled PSLTD for each block in a document with grid pooling.

\subsubsection{Using Correlation}
Pearson correlation coefficient, also referred to as Pearson's r, is a statistic that estimate the linear correlation between two variables as follows~\cite{swinscow2002statistics}.
\begin{equation}
r=\frac{\Sigma[(x-\bar{x})(y-\bar{y})]} {\sqrt{\Sigma[(x-\bar{x})^{2}(y-\bar{y})^{2}}]}
\end{equation}
Here, $x$ and $y$ represent the values of two variables.
$\Bar{x}$ and $\Bar{y}$ denote the mean of all observed values in $x$ and $y$, respectively.
The value of Pearson's r is between $+1$ and $-1$, where a value of $+1$ denotes a total positive linear correlation, $0$ is no linear correlation, and $-1$ is a total negative linear correlation.
We also analyze the effectiveness of features independent of a complex classifier.
For this, we calculate the Pearson correlation coefficient (termed as correlation value from here on) between pooled PSLTDs from train and test documents for each block.
A PSLTD pooled from a bock (i.e., a column or grid) in the test document is correlated with all PSLTDs pooled from that block in training documents printed by all the printers.
Note that there is only a single pooled PSLTD for each block in a document for grid pooling.
The printer corresponding to the largest correlation value between pooled PSLTDs in train and test is predicted as the pooled feature vector's source printer label.
We repeat this process for all pooled samples in the test.
Similar to the SVM method, a majority vote is taken over all predicted printer labels corresponding to groups of letters to obtain the printer label for the document under test.

%% file: Sections/S4_Expererimental.tex
\section{Experimental Evaluation}
\label{sec:Experimental}
We evaluate The performance of the proposed method on two datasets: a publicly available dataset~\cite{ferreira2015laser}, termed as DB1 from here on and another dataset~\cite{joshi2018single} containing documents printed by 18 printers in four font types (termed as DB2).
We do an extensive parameter search for both column and grid pooling to analyze printer signature variations.
We chose DB2 as our baseline dataset for this purpose as it allows analysis under cross font scenario.
The dataset DB2 is a higher precision dataset consisting of 16-bit depth intensity values compared to 8-bit depth in DB1.
So, we analyze the impact of the correlation-based prediction technique on the DB2 dataset.
The full-length PSLTD is a feature vector of 10502 dimensions~\cite{joshi2019source}.
We also compare the full-length feature vector's performance and the approximated feature vector of a smaller length.
In the remainder of this paper, we use PSLTD$_{4k}$ and PSLTD$_{10k}$ to denote the methods based on approximated smaller length feature vector and the original full-length feature vector, respectively.
Also, we use the general term PSLTD has been used to denote PSLTD$_{4k}$ to explain the experiments as we use PSLTD$_{4k}$ for all experiments except in Section~\ref{subsec:PSLTD_10k}.
The performance of our technique is compared with hand-crafted methods including
GLCM~\cite{Mikkilineni2005},
multi-directional GLCM \sloppy (\emph{GLCM\_MD})~\cite{ferreira2015laser}, multi-directional multi-scale GLCM (\emph{GLCM\_MD\_MS})~\cite{ferreira2015laser}, $\emph{CTGF-GLCM-MD-MS}_{e}$~\cite{ferreira2015laser} and \text{\emph{CC-RS-LTrP-PoEP}}~\cite{joshi2018single}.
We also compare with data-driven methods of \cite{ferreira2017data} denoted by \emph{CNN-}$\{S^{raw}, S^{med}, S^{avg}\}_{a,e}$ and~\cite{joshiIcassp2018} denoted as~\emph{CNN-}$\{S^{raw}, S^{nr}\}_{a,e}$.

\input{Tables/Dataset_DB2.tex}
\subsection{Dataset and Experimental Setup:} 
We used the publicly available dataset DB1~\cite{ferreira2015laser} and our dataset, DB2~\cite{joshi2018single}, to evaluate our method's performance.
DB1 contains 1184 Wikipedia pages (documents) in English and Portuguese language.
These pages are printed from 10 printers, including two of the same brand and same model.
This dataset consists of document images scanned at 600 dpi via the Plustek SO PL2546 scanner.
The image is available in an 8-bit format.
The documents comprise letters printed in mixed font types and sizes distributed randomly. 
Some documents contain bold and italic font styles also.

We created a dataset (Table ~\ref{tab:Dataset_DB2}) consisting of 720 pages printed from 18 laser and inkjet printers to examine the effect of font types.
These include three printers of the same brand and model.
The documents contain random text in the English language printed using four different font types.
However, in contrast with the DB1 dataset, a single document contains only a specific font type.
For each printer, there are twenty-five pages (documents) in Cambria (EC) font, while five pages each are in Arial (EA), Comic Sans (ES), and Times New Roman (ET) fonts. 
Pages in Arial font have font size 11, while pages in the other three fonts have font size 12 as per the general settings used in most legal documents. 
The dataset includes Cambria, Arial, and Times New Roman, as most legal documents widely use them.
In contrast, comic Sans is included because it looks considerably different from the rest of the fonts.
A single reference scanner (Epson Perfection V600 Photo Scanner) scanned all printed pages at 600 dpi and 300 dpi resolutions. 
The image is available in a 16-bit format. 
Three printers are of the same brand and model (LC9, LC10, and LC11).

For all the experiments, we fix the train and test sets in a disjoint manner.
The intensity and gradient threshold is kept fixed at $T_{0}$ = 20, $T_{1}$ = 80, $G_{0}$ = 90 for DB1 as it is scanned using 256 grayscale levels while $T_{0}$ and $T_{1}$ are set to 13000 and 50000, respectively for DB2 as it is scanned using 65536 grayscale levels.
All experiments have been performed using Matlab 2018b software.
The C - SVM with the radial basis function kernel of LIBSVM~\cite{Chang2013} is used for classification. Parameters (c and gamma) of SVM are chosen individually for each experiment using the search option available in LIBSVM with c $\in$ [-5, 15], and g $\in$ [-15, 3].
The step size of the grid search is fixed at 2.
The validation set is a subset of training data.
The experimental settings are consistent with the earlier works~\cite{joshi2018single,joshi2019source,ferreira2017data,joshiIcassp2018}.

\subsection{Parameter Search}
\input{Tables/DB2_col_search.tex}
\input{Tables/DB2_Letters_per_col.tex}
We conduct a complete parameter search for both column and grid pooling approaches of the proposed method.
DB2 allows the comparison under both the same and cross-font scenarios.
All experiments have been conducted using five pages containing text printed in Cambria font for training.
There are 20 pages in testing: Cambria font and 5 pages each of Arial, Times New Roman, and Comic Sans fonts.
All experimental results have been reported over five iterations, i.e., five unique and disjoint combinations of train and test set fixed for all parameter variations.

\subsubsection{Column Pooling}
For column pooling, we analyze the effect of a different number of columns (N$_{c}$) as well as the pooling parameter (N$_{p}$) on the classification accuracies.
Expressly, the value of N$_{c}$ is varied from 11 to 20 whereas, the value of N$_{p}$ is fixed at maximum possible value, i.e., all letters in a column are pooled into a single pooled feature vector (Table~\ref{tab:DB2_col_search}).
Preliminary experimental results with varying values of N$_{p}$ showed that the values do not make much of an impact as long as the number of samples is sufficient to train an SVM model.
We ensure this by choosing sufficiently smaller values of N$_{c}$ but large enough to capture the variation in printer signature.
The classification accuracy remains constant for the same font experiments.
For cross font experiments, the classification accuracy does not vary monotonically, and there is no single parameter pair that provides the highest classification accuracy for all three fonts.
For rest of the experiments, we chose N$_{c}$ = 15 and N$_{p}$ = `all' as this setting provides a significant number of samples (Table~\ref{tab:DB2_Letters_per_col}) to learn a separate classifier model for each column.
The average number of letters per column (using N$_{c}$ = 15) for all printers has been reported in Table~\ref{tab:DB2_Letters_per_col}.
The Table shows that the proposed column pooling strategy evenly distributes letters across all printers in our dataset.

\input{Tables/DB2_grid_search.tex}
\subsubsection{Grid Pooling}
For grid pooling, we analyze the effect of a different number of grids (N$_{w}$ $\times$ N$_{h}$) on the classification accuracies.
The number of grids is varied from 2 to 8 while also varying the number of grid blocks N$_{w}$ and N$_{h}$ with respect to each other (Table~\ref{tab:DB2_grid_search}).
The classification accuracy remains constant for all varying values of parameters.
The cross font accuracies with Arial font in general increases with the number of grid blocks.
The results with N$_{w}$ > 8 (not reported in the paper) showed that the average cross font accuracy does not increase significantly.
There is no specific trend observed in the other two fonts.
Like column pooling, there is no single parameter setting that provides the highest classification accuracy for all three fonts.
However, 8 $\times$ 8 grid configuration provides reasonably high accuracies for all three font types. So we chose this setting for the remaining experiments.

\subsection{Full Length PSLTD vs. Approximated PSLTD}
\label{subsec:PSLTD_10k}
\input{Tables/DB2_10k.tex}
The performance of proposed method using approximated PSLTD (i.e., PSLTD$_{4k}$) is compared with that using the full length PSLTD (i.e., PSLTD$_{10k}$) in Table~\ref{tab:DB2_10k}.
The approximated PSLTD of reduced dimensions performs similar to their full-length counterpart for both same and cross font experiments.
Thus to reduce complexity and save time, we use the approximated PSLTD$_{4k}$ as feature vectors with our proposed method for the remaining experiments.

\subsection{Block Pooling vs. Consecutive Pooling}
For comparison between proposed block pooling and consecutive pooling~\cite{joshi2018single}, we first visualize the feature vector in a reduced dimension using linear discriminant analysis (LDA). Secondly, Pearson's r correlation is used for comparison.
We choose the same training data (i.e., five printed documents per printer containing the same font type) of dataset DB2~\cite{joshi2018single} for both pooling methods.

\subsubsection{Comparison using LDA}
\begin{figure*}[htb!]
\centering
\begin{subfigure}{0.45\textwidth}
    \centering
    \includegraphics[width = \linewidth]{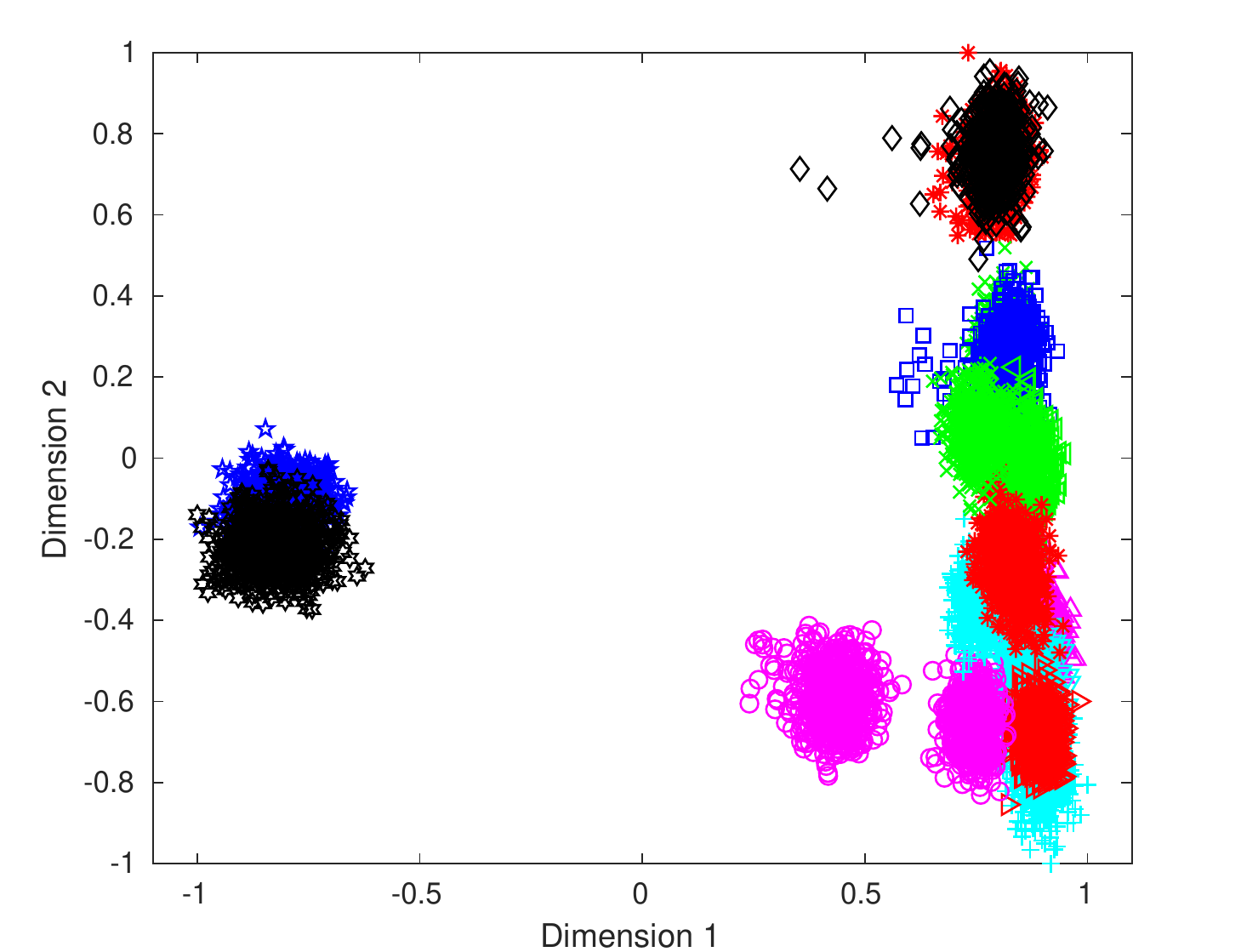}
    \caption{Consecutive pooling}\label{fig: Consecutive pooling}
\end{subfigure}
\begin{subfigure}{0.45\textwidth}
  \centering
  \includegraphics[width = \linewidth]{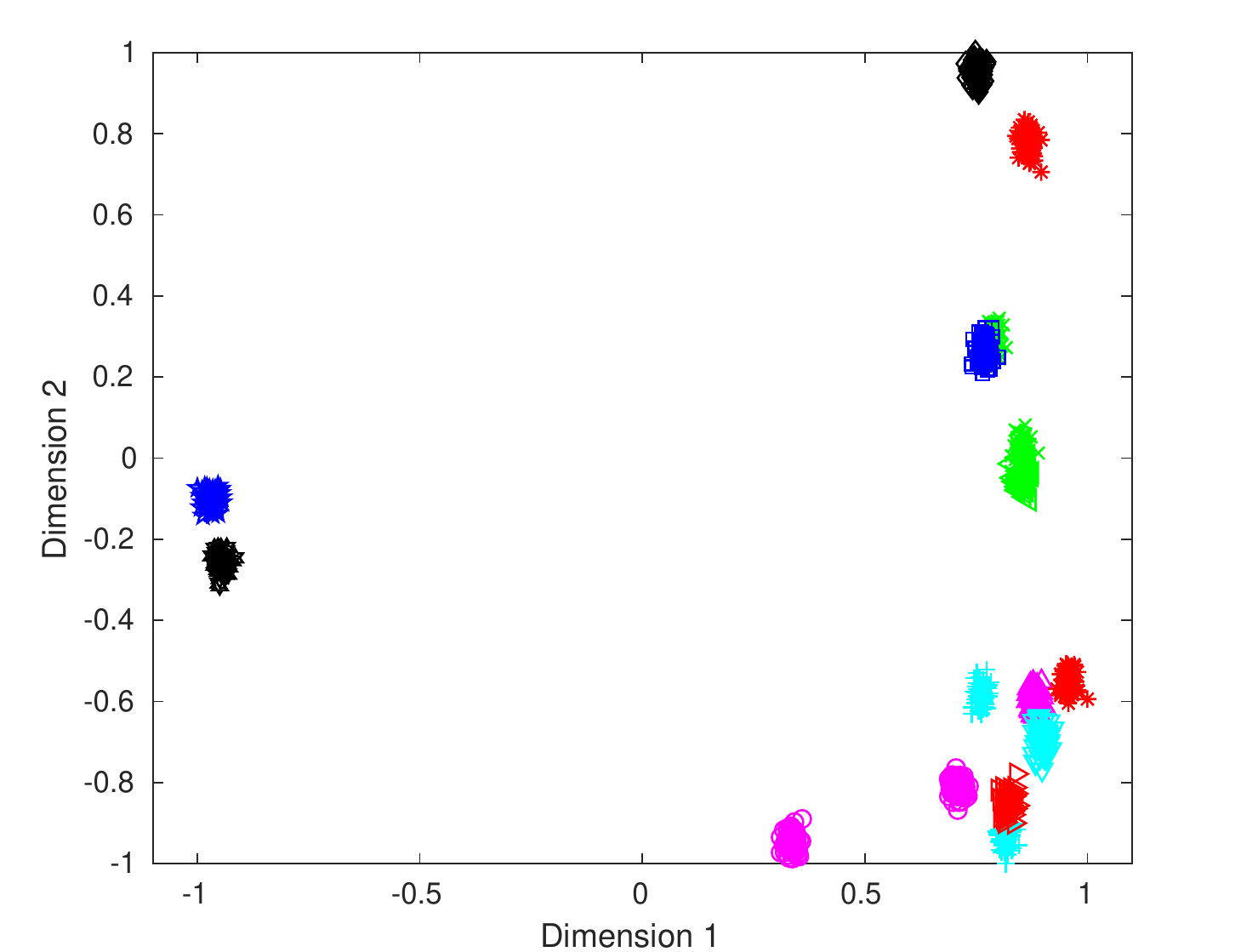}
  \caption{Column pooling.}\label{fig: column_wise_pooling1}
\end{subfigure}
\begin{subfigure}{0.50\textwidth}
  \centering
  \includegraphics[width = \linewidth]{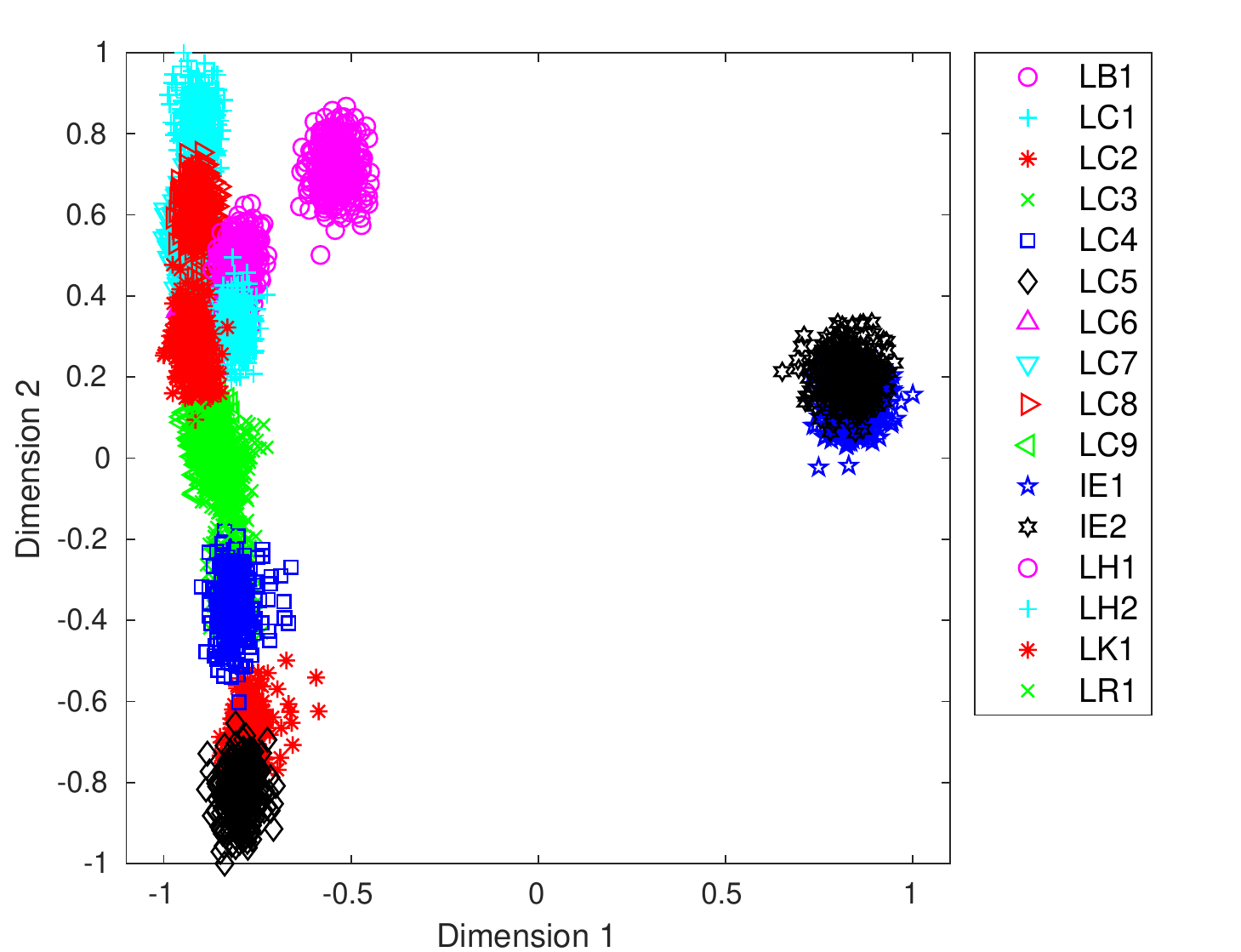}
  \caption{Grid pooling.}\label{fig: column_wise_pooling2}
\end{subfigure}
{\caption{Comparison of pooling techniques using first two components of LDA corresponding to maximum eigenvalues.}
\label{fig:lda}}
\end{figure*}

To visualize the distribution of features in a lower-dimensional space, we extracted PSLTDs from training samples of DB2.
We pool PSLTDs using N$_{p}$ equals 20 via consecutive pooling, column pooling for fifteen columns (i.e,, N$_{c}$ = 15) and grid pooling using 8 $\times$ 8 grid.
Then we apply linear discriminant analysis (LDA) to reduce the feature's dimension.
The first two dimensions corresponding to the largest eigenvalues of projected data are plotted in figure~\ref{fig:lda}.
These plots indicate that both column and grid pooling method reduces intra-class separation and increases inter-class separation.
The figure depicts that different clusters corresponding to various printers have more discrimination in column pooling than consecutive pooling.
As expected, the improvement in class-wise cluster separation is drastically better with column pooling than grid pooling.
It provides visual clues that column pooling offers us a better representation with less overlapping of different class samples.

\input{Tables/DB2_corr_comparison.tex}
\subsubsection{Comparison using Correlation} 
Following our method's primary hypothesis, we consider that sample from the same column of a printer's pages has more similarity than cross columns.
We compute the correlation values between pooled PSLTDs using consecutive pooling, column pooling, and grid pooling to observe this phenomenon.
For consecutive pooling, the correlation values are calculated between all pooled PSLTDs of a printer for all 18 printers in DB2.
On the other hand, for column and grid pooling, correlation values are calculated between pooled PSLTDs belonging to a printer's training documents but belonging to the same block (column or grid).
The median of correlation values obtained by the same block correlation is consistently higher than that of consecutive pooling for all printers (Table~\ref{tab:DB2_corr_comparison}).
This observation establishes the validity of our primary hypothesis about the variation of printer signature across a document image and its effectiveness compared to consecutive pooling.
However, the difference between correlation values obtained using consecutive and block pooling is seen only after 4$^{th}$ or 5$^{th}$ digit after the decimal point.
So, the document images need to be scanned using a high precision of 16-bit depth images.

\subsection{Comparison with State-of-the-art}
\input{Tables/DB1_comparison_e.tex}
\input{Tables/DB1_comparison_all.tex}
We compare the performance of the proposed method with existing state-of-the-art methods using a combination of datasets DB1 and DB2. For DB1, the proposed method's efficacy is analyzed using an ensemble of SVMs, while the higher precision of DB2 allows us to use correlation-based prediction.
\subsubsection{Experiment on Dataset DB1} 
First, we evaluated the performance of our method on dataset DB1. 
As only 8-bit document images available in DB1, we use our SVM-based approach.
We analyzed the performance using only letter type `e' (the most frequently occurring letter in the English language) and all connected components printed on a document.
We use the same train and test folds used in previous works~\cite{ferreira2015laser,ferreira2017data,joshi2018single,joshi2019source} for their result based on a 5 $\times$ 2 cross-validation method for a fair comparison.
Each fold has around 592 pages each for training and testing.
The performance of the proposed method has been compared with various state-of-the-art methods using all occurrences of letter type `e'.
This approach allows a fair comparison consistently with many other baseline methods developed for a specific letter type.
The proposed method achieves an accuracy of 97.76\% and 92.93\% using column (N$_{c}$=15) and grid (8 $\times$ 8) pooling, respectively (Table~\ref{tab:DB1_comparison_e}).
Our proposed method using column pooling is denoted by Proposed$_{col5}$ and Proposed$_{col15}$ for N$_{c}$=5 and 15, respectively.
Whereas, our proposed grid pooling-based method is denoted by Proposed$_{grid}$.
Results show that grid pooling does not perform well.
A possible reason could be that the number of samples per SVM is insufficient to learn a discriminative model.
Note that the proposed method using 8 $\times$ 8 grid pooling has an ensemble of 64 SVMS.
We also try other pooling parameters and find that the proposed method performs slightly better with $N_{c}$=5 as compared to consecutive pooling.
PSLTD$_{e}$ and proposed method (PSLTD$_{4k,e}$) outperforms all other existing methods.
We further compare PSLTD$_{e}$ and proposed method using all printed letters as reported in Table~\ref{tab:DB1_comparison_all}.
We observe that the proposed method performs similarly to the PSLTD method.
Nonetheless, the font type and size characteristics of train and test data in DB1 are not well defined.
So, we also carry out experiments on the DB2 dataset, which allows analysis in an ideal cross-font scenario.

\subsection{Experiments on dataset DB2}
We evaluated the performance of the proposed method on DB2 (having approximately 2300 connected components per document) for (a) inter-model scenario (i.e., no two printers are of the same brand and model) and (b) intra-model scenario (i.e., multiple printers of same brand and model).
We extract all connected components for experiments on DB2.
Dataset DB2 consists of documents scanned at 16 - bit (65536 - scale). 
So, unlike DB1, on DB2 we use correlation-based prediction in our proposed method.
\input{Tables/DB2_comparison_20Pages.tex}
\subsubsection{Intermodel scenario}
For the inter-model scenario, we analyzed the proposed method's performance on 16 printers (except LC10, LC11 in table \ref{tab:Dataset_DB2}). 
There are 20 Cambria font documents per printer in training and five Cambria font documents in testing.
We also test on all five documents containing text printed in Arial (A), Times New Roman (T), and Comic Sans font (S) types.
The results are reported over five iterations, i.e., five unique combinations of train and test data.
In this setting, PSLTD~\cite{joshi2019source} and column pooling variants of proposed method achieve 100\% classification accuracy in the same font scenario as reported in Table~\ref{tab:DB2_comparison_20Pages}.
The grid pooling variant of the proposed method achieves 99.50\% accuracy.
Note that, in original PSLTD~\cite{joshi2019source}, we need to learn an SVM model, but the proposed method predicts using a simple correlation-based prediction technique without the requirement of learning a complex classifier model.
Moreover, the proposed method surpasses the performance of state-of-the-art PSLTD-based method~\cite{joshi2019source} under the cross-font scenario.
Specifically, the grid pooling variant (i.e., 8$\times$8 grid) of the proposed method achieves average classification accuracies of 93.50\%, 94.25\%, and 60.25\% when trained using documents containing Cambria font and tested using documents containing A, T, and S font types, respectively.
The column pooling variant (i.e., N$_{c}$=15) of the proposed method achieves average classification accuracies of 85.25\%, 96.25\%, and 54.75\%, respectively with A, T, and S.
The results show that neither variant of the proposed method is consistently better than the other on all font types.
However, it can be seen that the overall performance of grid column pooling is better than column pooling under the cross-font scenario on DB2.

\input{Tables/DB2_samebrandmodel.tex}
\subsubsection{Intra-model scenario}  
The intra-model scenario analyzes the performance of the proposed method on printers of the same brand and model.
For the Intra model scenario, DB2 consists of three printers of the same brand and model.
We choose the same folds of 20 and 5 documents in training and testing, respectively, over five iterations. Under these settings, 
The proposed method achieves 100\% classification accuracy, as shown in Table~\ref{tab:DB2_samebrandmodel} whereas, there is some confusion between printers using PSLTD-based method~\cite{joshi2019source}.
The proposed method correctly classifies documents printed from printers of the same brand and model without learning a complex classifier.
We further analyze the proposed method's performance using all 18 printers of DB2, achieving an average classification accuracy of 100\% under the same font scenario for both column and grid polling variants.
The experimental setting remains the same as in the inter-model scenario, i.e., 20 documents of Cambria in train and rest five documents in the test.
The column pooling methods achieve 84.89\%, 91.33\%, and 52.67\% accuracies under the cross-font scenario, and grid pooling achieves 92.67\%, 89.11\%, and 56.89\%, respectively on A, T, and S font types.
Similar to the intra-model scenario, grid pooling performs better than the inter-model scenario.
The confusion matrices corresponding to grid pooling have been depicted in Tables~\ref{tab:DB2_ConfMat_18P_AT} and \ref{tab:DB2_ConfMat_18P_S}.
For font type A, the proposed method correctly classifies 15 out of 18 printers for all test documents, whereas, for font type T, all documents printed by 16 out of 18 printers are correctly classified.
As expected, there is some confusion between the printer of the same brand and model for font types A and T.
There is a lot of confusion among many printers for font type S.
The results show that block pooling achieves significant performance under cross font scenarios when font types in train and test are not drastically different.
The discriminative power of block pooling is highlighted by a simple correlation-based prediction instead of a complex classifier.
\input{Tables/DB2_ConfMat_18P_AT.tex}
\input{Tables/DB2_ConfMat_18P_S.tex}

%% file: Tables/Dataset_DB2.tex
\begin{table}[htb!]
\centering
\caption{Details of dataset DB2~\cite{joshi2018single}).}
\label{tab:Dataset_DB2}
\resizebox{\textwidth}{!}{
\begin{tabular}{|c|c|c|c|c|c|}
\hline
\textbf{S. No.} & \textbf{\begin{tabular}[c]{@{}c@{}}Printer \\ ID\end{tabular}} & \textbf{\begin{tabular}[c]{@{}c@{}}Printer \\ Brand\end{tabular}} & \textbf{\begin{tabular}[c]{@{}c@{}}Printer \\ Model\end{tabular}} & \textbf{\begin{tabular}[c]{@{}c@{}}Printer Resolution \\ (in dpi)\end{tabular}} & \textbf{\begin{tabular}[c]{@{}c@{}}Printer \\ Type\end{tabular}} \\ \hline
1               & LB1                                                            & Brothers                                                          & DCP 7065DN                                                        & $2400 \times 600$                                                               & Laser                                                            \\ \hline
2               & LC1                                                            & Canon                                                             & D520                                                              & $1200 \times 600$                                                               & Laser                                                            \\ \hline
3               & LC2                                                            & Canon                                                             & 16570                                                             & $2400 \times 600$                                                               & Laser                                                            \\ \hline
4               & LC3                                                            & Canon                                                             & IR 5000                                                           & $2400 \times 600$                                                               & Laser                                                            \\ \hline
5               & LC4                                                            & Canon                                                             & IR 7095                                                           & $1200 \times 600$                                                               & Laser                                                            \\ \hline
6               & LC5                                                            & Canon                                                             & IR 8500                                                           & $2400 \times 600$                                                               & Laser                                                            \\ \hline
7               & LC6                                                            & Canon                                                             & LBP 2900B                                                         & $2400 \times 600$                                                               & Laser                                                            \\ \hline
8               & LC7                                                            & Canon                                                             & LBP 5050                                                          & $9600 \times 600$                                                               & Laser                                                            \\ \hline
9               & LC8                                                            & Canon                                                             & MF 4320                                                           & $600 \times 600$                                                                & Laser                                                            \\ \hline
10              & LC9                                                            & \textbf{Canon}                                                    & \textbf{MF 4820d}                                                 & $600 \times 600$                                                                & Laser                                                            \\ \hline
11              & LC10                                                           & \textbf{Canon}                                                    & \textbf{MF 4820d}                                                 & $600 \times 600$                                                                & Laser                                                            \\ \hline
12              & LC11                                                           & \textbf{Canon}                                                    & \textbf{MF 4820d}                                                 & $600 \times 600$                                                                & Laser                                                            \\ \hline
13              & IE1                                                            & Epson                                                             & L800                                                              & $5760 \times 1440$                                                              & Inkjet                                                           \\ \hline
14              & IE2                                                            & Epson                                                             & EL 360                                                            & $1200 \times 600$                                                               & Inkjet                                                           \\ \hline
15              & LH1                                                            & HP                                                                & 1020                                                              & $600 \times 600$                                                                & Laser                                                            \\ \hline
16              & LH2                                                            & HP                                                                & M1005                                                             & $600 \times 600$                                                                & Laser                                                            \\ \hline
17              & LK1                                                            & Konica Minolta                                                    & Bizhub 215                                                        & $600 \times 600$                                                                & Laser                                                            \\ \hline
18              & LR1                                                            & Ricoh                                                             & MP 5002                                                           & $600 \times 600$                                                                & Laser                                                            \\ \hline
\end{tabular}
}
\end{table}

%% file: Tables/DB2_col_search.tex
\begin{table}[htb!]
\centering
\caption{Parameter searching for column pooling. Average classification accuracies (in \%) on 20 documents of Cambria (C) font and 5 documents each of Arial (A), Times New Roman (T), and Comic Sans (S) fonts using 5 documents of C font in training.}
\label{tab:DB2_col_search}
\begin{tabular}{|c|c|c|c|c|c|}
\hline
\textbf{\begin{tabular}[c]{@{}c@{}}No. of \\ Columns\\ (N$_{c}$)\end{tabular}} & \textbf{\begin{tabular}[c]{@{}c@{}}Pooling\\ Parameter\\ (N$_{p}$)\end{tabular}} & \textbf{C} & \textbf{A}   & \textbf{T}   & \textbf{S}    \\ \hline
\textbf{11}                                                               & \textbf{all}                                                                & 98.94       & 78            & \textbf{91.5} & 47.5           \\ \hline
\textbf{12}                                                               & \textbf{all}                                                                & 98.94       & 79.5          & 90.25         & \textbf{51.75} \\ \hline
\textbf{13}                                                               & \textbf{all}                                                                & 98.94       & 81.5          & 90.75         & 48.5           \\ \hline
\textbf{14}                                                               & \textbf{all}                                                                & 98.94       & 80.25         & 89.75         & 48.5           \\ \hline
\textbf{15}                                                               & \textbf{all}                                                                & 98.94       & 80.25         & \textbf{91.5} & 49.25          \\ \hline
\textbf{16}                                                               & \textbf{all}                                                                & 98.94       & 81.25         & 89.75         & 48.5           \\ \hline
\textbf{17}                                                               & \textbf{all}                                                                & 98.94       & 79.5          & 89.5          & 49             \\ \hline
\textbf{18}                                                               & \textbf{all}                                                                & 98.94       & \textbf{83.5} & 89.75         & 48.75          \\ \hline
\textbf{19}                                                               & \textbf{all}                                                                & 98.94       & \textbf{83.5} & 89.5          & 47.5           \\ \hline
\textbf{20}                                                               & \textbf{all}                                                                & 98.94       & 80.25         & 89.25         & 47.75          \\ \hline
\end{tabular}
\end{table}

%% file: Tables/DB2_Letters_per_col.tex
\begin{table*}[htb!]
\centering
\caption{Average letters per column (N$_{c}$=15) for all printers in dataset DB2.}
\label{tab:DB2_Letters_per_col}
\resizebox{\textwidth}{!}{
\begin{tabular}{|c|c|c|c|c|c|c|c|c|c|c|c|c|c|c|c|}
\hline
\textbf{Printer ID} & \textbf{1} & \textbf{2} & \textbf{3} & \textbf{4} & \textbf{5} & \textbf{6} & \textbf{7} & \textbf{8} & \textbf{9} & \textbf{10} & \textbf{11} & \textbf{12} & \textbf{13} & \textbf{14} & \textbf{15} \\ \hline
\textbf{LB1}        & 190        & 177        & 175        & 170        & 167        & 175        & 172        & 172        & 168        & 169         & 167         & 167         & 169         & 162         & 185         \\ \hline
\textbf{LC1}        & 188        & 179        & 174        & 170        & 168        & 172        & 172        & 173        & 167        & 165         & 166         & 166         & 166         & 159         & 179         \\ \hline
\textbf{LC2}        & 191        & 175        & 174        & 170        & 168        & 174        & 173        & 170        & 167        & 167         & 166         & 165         & 166         & 158         & 184         \\ \hline
\textbf{LC3}        & 189        & 177        & 175        & 170        & 168        & 175        & 172        & 172        & 167        & 168         & 168         & 167         & 169         & 161         & 187         \\ \hline
\textbf{LC4}        & 189        & 178        & 173        & 170        & 169        & 172        & 171        & 171        & 167        & 166         & 167         & 165         & 168         & 159         & 184         \\ \hline
\textbf{LC5}        & 166        & 152        & 152        & 149        & 147        & 151        & 151        & 150        & 149        & 144         & 146         & 147         & 147         & 139         & 162         \\ \hline
\textbf{LC6}        & 190        & 179        & 175        & 169        & 168        & 173        & 171        & 172        & 169        & 168         & 166         & 166         & 168         & 163         & 187         \\ \hline
\textbf{LC7}        & 190        & 178        & 175        & 171        & 168        & 174        & 172        & 173        & 169        & 168         & 167         & 167         & 168         & 161         & 187         \\ \hline
\textbf{LC8}        & 189        & 179        & 175        & 170        & 167        & 174        & 172        & 173        & 169        & 168         & 165         & 166         & 167         & 163         & 186         \\ \hline
\textbf{LC9}        & 173        & 170        & 169        & 163        & 162        & 166        & 162        & 159        & 155        & 150         & 148         & 147         & 141         & 140         & 161         \\ \hline
\textbf{LC10}       & 165        & 162        & 157        & 153        & 159        & 159        & 156        & 156        & 152        & 152         & 152         & 149         & 152         & 146         & 165         \\ \hline
\textbf{LC11}       & 186        & 177        & 171        & 169        & 164        & 168        & 166        & 167        & 161        & 158         & 161         & 156         & 157         & 149         & 171         \\ \hline
\textbf{IE1}        & 173        & 164        & 164        & 160        & 156        & 163        & 158        & 159        & 154        & 157         & 154         & 157         & 157         & 151         & 172         \\ \hline
\textbf{IE2}        & 187        & 175        & 173        & 170        & 166        & 172        & 171        & 170        & 167        & 166         & 166         & 165         & 168         & 161         & 185         \\ \hline
\textbf{LH1}        & 190        & 180        & 174        & 170        & 167        & 174        & 173        & 174        & 168        & 168         & 166         & 166         & 167         & 163         & 187         \\ \hline
\textbf{LH2}        & 190        & 180        & 175        & 169        & 167        & 173        & 174        & 173        & 168        & 167         & 166         & 166         & 167         & 164         & 187         \\ \hline
\textbf{LK1}        & 190        & 177        & 174        & 172        & 167        & 174        & 173        & 172        & 169        & 168         & 167         & 166         & 169         & 162         & 186         \\ \hline
\textbf{LR1}        & 190        & 177        & 176        & 171        & 168        & 175        & 172        & 173        & 167        & 169         & 167         & 166         & 169         & 161         & 186         \\ \hline
\end{tabular}
}
\end{table*}

%% file: Tables/DB2_grid_search.tex
\begin{table}[htb!]
\centering
\caption{Parameter searching for grid pooling. Average classification accuracies (in \%) on 20 documents of Cambria (C) font and 5 documents each of Arial (A), Times New Roman (T), and Comic Sans (S) fonts using 5 documents of C font in training.}
\label{tab:DB2_grid_search}
\begin{tabular}{|c|c|c|c|c|c|}
\hline
\textbf{N$_{w}$} & \textbf{N$_{h}$} & \textbf{C} & \textbf{A}    & \textbf{T}    & \textbf{S}    \\ \hline \hline
\textbf{2}       & \textbf{2}       & 98.94       & 72.50          & 90.50          & 52.75          \\ \hline
\textbf{2}       & \textbf{4}       & 98.94       & 77.25          & 90.75          & \textbf{53.00} \\ \hline
\textbf{2}       & \textbf{6}       & 98.94       & 78.50          & 91.75          & 50.75          \\ \hline
\textbf{2}       & \textbf{8}       & 98.94       & 80.25          & 91.00          & 49.75          \\ \hline
\textbf{4}       & \textbf{2}       & 98.94       & 78.25          & 89.75          & 50.75          \\ \hline
\textbf{4}       & \textbf{4}       & 98.94       & 82.75          & 92.00          & 51.25          \\ \hline
\textbf{4}       & \textbf{6}       & 98.94       & 85.75          & \textbf{92.75} & 52.25          \\ \hline
\textbf{4}       & \textbf{8}       & 98.94       & 86.00          & 91.00          & 50.75          \\ \hline
\textbf{6}       & \textbf{2}       & 98.94       & 82.75          & 90.00          & 48.75          \\ \hline
\textbf{6}       & \textbf{4}       & 98.94       & 87.75          & 92.50          & 50.25          \\ \hline
\textbf{6}       & \textbf{6}       & 98.94       & 87.00          & \textbf{92.75} & 51.00          \\ \hline
\textbf{6}       & \textbf{8}       & 98.94       & 88.75          & 92.00          & 49.50          \\ \hline
\textbf{8}       & \textbf{2}       & 98.94       & 83.00          & 90.00          & 50.00          \\ \hline
\textbf{8}       & \textbf{4}       & 98.94       & 89.50          & 92.25          & 49.50          \\ \hline
\textbf{8}       & \textbf{6}       & 98.94       & 89.50          & 91.25          & 52.75          \\ \hline
\textbf{8}       & \textbf{8}       & 98.94       & \textbf{90.00} & 92.25          & 49.50          \\ \hline
\end{tabular}
\end{table}

%% file: Tables/DB2_10k.tex
\begin{table}[htb!]
\centering
\caption{Comparison of 10k and 4k feature vectors. Average classification accuracies (in \%) on 20 documents of Cambria (C) font and 5 documents each of Arial (A), Times New Roman (T), and Comic Sans (S) fonts using 5 documents of C font in training.}
\label{tab:DB2_10k}
\resizebox{\textwidth}{!}{
\begin{tabular}{|c|c|c|c|c|c|c|c|c|c|}
\hline
\multicolumn{2}{|c|}{\textbf{}}                  & \multicolumn{2}{c|}{{\color[HTML]{00009B} \textbf{C}}}                         & \multicolumn{2}{c|}{{\color[HTML]{036400} \textbf{A}}}                       & \multicolumn{2}{c|}{{\color[HTML]{9A0000} \textbf{T}}}                       & \multicolumn{2}{c|}{\textbf{S}}                     \\ \hline
\textbf{N$_{c}$}             & \textbf{N$_{p}$}            & {\color[HTML]{00009B} \textbf{10k}}     & {\color[HTML]{00009B} \textbf{4k}}    & {\color[HTML]{036400} \textbf{10k}}   & {\color[HTML]{036400} \textbf{4k}}    & {\color[HTML]{9A0000} \textbf{10k}}   & {\color[HTML]{9A0000} \textbf{4k}}    & {\color[HTML]{000000} \textbf{10k}} & \textbf{4k}    \\ \hline
\textbf{15}             & \textbf{all}             & {\color[HTML]{00009B} 98.94}          & {\color[HTML]{00009B} 98.94} & {\color[HTML]{036400} 78.00}  & {\color[HTML]{036400} 80.25}          & {\color[HTML]{9A0000} \textbf{90.25}} & {\color[HTML]{9A0000} 91.50}          & \textbf{50.00}                             & 49.25 \\ \hline
\multicolumn{2}{|c|}{\textbf{Grid 8 $\times$ 8}} & {\color[HTML]{00009B} 98.94}          & {\color[HTML]{00009B} 98.94}               & {\color[HTML]{036400} 84.75}          & {\color[HTML]{036400} \textbf{90.00}}               & {\color[HTML]{9A0000} 91.00}             & {\color[HTML]{9A0000} \textbf{92.25}}               & \textbf{50.25}                               & 49.50               \\ \hline
\end{tabular}
}
\end{table}

%% file: Tables/DB2_corr_comparison.tex
\begin{table}[htb!]
\caption{Comparison of pooling techniques using median correlation values.}
\label{tab:DB2_corr_comparison}
\begin{tabular}{|c|c|c|c|}
\hline
\textbf{Printer ID} & \textbf{Consecutive Pooling}  & \textbf{Column Pooling}       & \textbf{Grid Pooling}   \\ \hline
\hline
\textbf{LB1}        & 0.9999403786 & 0.9999894087          & \textbf{0.9999932251} \\ \hline
\textbf{LC1}        & 0.9999417235 & \textbf{0.9999939426} & 0.9999928397          \\ \hline
\textbf{LC2}        & 0.9999496627 & \textbf{0.9999955784} & 0.9999918113          \\ \hline
\textbf{LC3}        & 0.9999536711 & 0.9999932549          & \textbf{0.9999933009} \\ \hline
\textbf{LC4}        & 0.9999521655 & \textbf{0.9999940128} & 0.9999921736          \\ \hline
\textbf{LC5}        & 0.9999686467 & \textbf{0.9999946214} & 0.9999937096          \\ \hline
\textbf{LC6}        & 0.9999703351 & 0.9999935464          & \textbf{0.9999952785} \\ \hline
\textbf{LC7}        & 0.9999716916 & \textbf{0.9999943718} & 0.9999941307          \\ \hline
\textbf{LC8}        & 0.9999756106 & 0.9999937039          & \textbf{0.9999948721} \\ \hline
\textbf{LC9}        & 0.9999789123 & 0.9999922834          & \textbf{0.99999445}   \\ \hline
\textbf{LC10}       & 0.9999812282 & \textbf{0.9999958219} & 0.9999927177          \\ \hline
\textbf{LC11}       & 0.9999814061 & 0.9999925254          & \textbf{0.9999929945} \\ \hline
\textbf{IE1}        & 0.9999829966 & \textbf{0.9999954384} & 0.9999886172          \\ \hline
\textbf{IE2}        & 0.9999834864 & \textbf{0.9999887612} & 0.9999882253          \\ \hline
\textbf{LH1}        & 0.9999884284 & 0.9999902606          & \textbf{0.9999933383} \\ \hline
\textbf{LH2}        & 0.9999898499 & \textbf{0.9999950545} & 0.9999945381          \\ \hline
\textbf{LK1}        & 0.999989456  & \textbf{0.9999945135} & 0.999990502           \\ \hline
\textbf{LR1}        & 0.9999909281 & 0.999991202           & \textbf{0.9999940467} \\ \hline
\end{tabular}
\end{table}

%% file: Tables/DB1_comparison_e.tex
\begin{table}[htb!]
\centering
\caption{Comparison of average page-level classification accuracies of the proposed method against state-of-the-art methods on dataset DB1~\cite{ferreira2015laser} using letter type `e' and $5 \times 2$ cross-validation on the train and test folds provided by~\cite{ferreira2017data}.}
\label{tab:DB1_comparison_e}
\begin{tabular}{|l|c|}
\hline
\multicolumn{1}{|c|}{\textbf{Method}}                                                        & \textbf{\begin{tabular}[c]{@{}c@{}}Accuracy\\ (in \%)\end{tabular}}
\\ \hline
\hline
\textbf{\emph{GLCM}$_{e}$}~\cite{Mikkilineni2005}                               & 77.87                                                               \\ \hline
\textbf{\emph{GLCM\_MD}$_{e}$}~\cite{ferreira2015laser}                         & 91.08 
               \\ \hline
\textbf{\emph{GLCM\_MD\_MS}$_{e}$}~\cite{ferreira2015laser}                     & 94.30                        \\ \hline
\textbf{\emph{CTGF-GLCM-MD-MS}$_{e}$}~\cite{ferreira2017data}                   & 96.26                       \\ \hline
\textbf{\emph{CNN-}$\{S^{raw}\}_{e}$}~\cite{ferreira2017data}   & 96.13 \\ \hline
 \textbf{\emph{CNN-}$\{S^{raw}, S^{med}, S^{avg}\}_{a,e}$}~\cite{ferreira2017data} & 97.33  \\ \hline
\textbf{\text{\emph{CC-RS-LTrP-PoEP}}}$_{e}$~\cite{joshi2018single} &  97.12    \\ \hline
\textbf{PSLTD$_{e}$~\cite{joshi2019source}}                                                                 & \textbf{98.92}                    \\ \hline
\textbf{Proposed$_{col5,e}$}                                                                 & 98.61                    \\ \hline
\textbf{Proposed$_{col15,e}$}                                                                 & 97.76                    \\ \hline
\textbf{Proposed$_{grid,e}$}                                                                 & 92.93                    \\ \hline
\end{tabular}
\end{table}

%% file: Tables/DB1_comparison_all.tex
\begin{table}[htb!]
\centering
\caption{Comparison of average page-level classification accuracies of the proposed method against PSLTD on dataset DB1~\cite{ferreira2015laser} using all letter types and $5 \times 2$ cross-validation on the train and test folds provided by~\cite{ferreira2017data}.}
\label{tab:DB1_comparison_all}
\begin{tabular}{|l|c|}
\hline
\hline
\multicolumn{1}{|c|}{\textbf{Method}}                                                        & \textbf{\begin{tabular}[c]{@{}c@{}}Accuracy\\ (in \%)\end{tabular}}
\\ \hline
\textbf{PSLTD~\cite{joshi2019source}}                                                                 & \textbf99.27                    \\ \hline
\textbf{Proposed$_{col5}$}                                                                 & \textbf{99.37}                    \\ \hline
\textbf{Proposed$_{col15}$}                                                                 & 99.16                    \\ \hline
\textbf{Proposed$_{grid}$}                                                                 & 96.77                    \\ \hline
\end{tabular}
\end{table}

%% file: Tables/DB2_comparison_20Pages.tex
\begin{table}[htb!]
	\centering
	\caption{Average classification accuracies (in \%) for same and cross font experiments using sixteen printers of unique brand and model in DB2~\cite{joshi2018single} (i.e., except LC10 and LC11).}
	\label{tab:DB2_comparison_20Pages}
	\begin{tabular}{|c|l|c|c|c|}
		\hline
		\textbf{Train Font} & \multicolumn{4}{c|}{\textbf{Cambria (C)}}                                                          \\ \hline
		\textbf{Test Font}  & \textbf{C}   & \textbf{A}                 & \textbf{T}                 & \textbf{S}                \\ \hline
		\textbf{PSLTD~\cite{joshi2019source}}   & \textbf{100}        & 80.00             & \textbf{98.75}             & 54.25 
		\\ \hline
		\textbf{Proposed$_{col15}$}   & \textbf{100}        & 85.25             & 96.25             & 54.75            \\ \hline
		\textbf{Proposed$_{grid}$}   & 99.50        & \textbf{93.50}             & 94.25             & \textbf{60.25}            \\ \hline
	\end{tabular}
\end{table}

%% file: Tables/DB2_samebrandmodel.tex
\begin{table}[htb!]
    \centering
    \caption{Comparison of mean confusion matrices (accuracy in \%) for classifying 3 printers of same brand and model. Bold values correspond to our proposed method while values in bracket correspond to PSLTD using consecutive pooling~\cite{joshi2019source}.}
    \label{tab:DB2_samebrandmodel}
    \begin{tabular}{|c|c|c|c|}
        \hline
        \multicolumn{1}{|l|}{\textbf{\begin{tabular}[c]{@{}l@{}}Predicted Class\\ True Class\end{tabular}}} & \textbf{LC9}   & \textbf{LC10}   & \textbf{LC11}   \\ \hline
        \textbf{LC9}                                                                                        & \textbf{100} {[}\textbf{100}{]} & 0     & 0     \\ \hline
        \textbf{LC10}                                                                                       & 0    & \textbf{100} {[}97.1{]} & \textbf{0} {[}2.9{]}   \\ \hline
        \textbf{LC11}                                                                                       & 0      & \textbf{0} {[}1.2{]}   & \textbf{100} {[}98.8{]} \\ \hline
    \end{tabular}
\end{table}

%% file: Tables/DB2_ConfMat_18P_AT.tex
\begin{table}[htb!]
\caption{Mean confusion matrices (in \%) for Arial and Times New Roman font types under cross font scenario using all printers in DB2\cite{joshi2018single} obtained with grid pooling and correlation-based prediction.}
\label{tab:DB2_ConfMat_18P_AT}
\resizebox{\textwidth}{!}{
\begin{tabular}{|c|c|c|c|c|c|c|c|c|c|c|c|c|c|c|c|c|c|c|}
\hline
\multicolumn{19}{|c|}{\textbf{Arial (A)}}                                                                                                                                                                                                                                                                                                                                                                                                                                                                                                                                                                                                                                                                                                                    \\ \hline
\multicolumn{1}{|l|}{\textbf{}} & \textbf{LB1}                         & \textbf{LC1}                         & \textbf{LC2}                         & \textbf{LC3}                         & \textbf{LC4}                         & \textbf{LC5}                         & \textbf{LC6}                         & \textbf{LC7}                         & \textbf{LC8}                         & \textbf{LC9}                         & \textbf{LC10}                        & \textbf{LC11}                       & \textbf{IE1}                         & \textbf{IE2}                         & \textbf{LH1}                         & \textbf{LH2}                         & \textbf{LK1}                         & \textbf{LR1}                         \\ \hline
\textbf{LB1}                    & \cellcolor[HTML]{9AFF99}\textbf{100} & 0                                    & 0                                    & 0                                    & 0                                    & 0                                    & 0                                    & 0                                    & 0                                    & 0                                    & 0                                    & 0                                   & 0                                    & 0                                    & 0                                    & 0                                    & 0                                    & 0                                    \\ \hline
\textbf{LC1}                    & 0                                    & \cellcolor[HTML]{9AFF99}\textbf{100} & 0                                    & 0                                    & 0                                    & 0                                    & 0                                    & 0                                    & 0                                    & 0                                    & 0                                    & 0                                   & 0                                    & 0                                    & 0                                    & 0                                    & 0                                    & 0                                    \\ \hline
\textbf{LC2}                    & 0                                    & 0                                    & \cellcolor[HTML]{9AFF99}\textbf{100} & 0                                    & 0                                    & 0                                    & 0                                    & 0                                    & 0                                    & 0                                    & 0                                    & 0                                   & 0                                    & 0                                    & 0                                    & 0                                    & 0                                    & 0                                    \\ \hline
\textbf{LC3}                    & 0                                    & 0                                    & 0                                    & \cellcolor[HTML]{9AFF99}\textbf{100} & 0                                    & 0                                    & 0                                    & 0                                    & 0                                    & 0                                    & 0                                    & 0                                   & 0                                    & 0                                    & 0                                    & 0                                    & 0                                    & 0                                    \\ \hline
\textbf{LC4}                    & 0                                    & 0                                    & 0                                    & \cellcolor[HTML]{FFFC9E}4            & \cellcolor[HTML]{9AFF99}\textbf{96}  & 0                                    & 0                                    & 0                                    & 0                                    & 0                                    & 0                                    & 0                                   & 0                                    & 0                                    & 0                                    & 0                                    & 0                                    & 0                                    \\ \hline
\textbf{LC5}                    & 0                                    & 0                                    & 0                                    & 0                                    & 0                                    & \cellcolor[HTML]{9AFF99}\textbf{100} & 0                                    & 0                                    & 0                                    & 0                                    & 0                                    & 0                                   & 0                                    & 0                                    & 0                                    & 0                                    & 0                                    & 0                                    \\ \hline
\textbf{LC6}                    & 0                                    & 0                                    & 0                                    & 0                                    & 0                                    & 0                                    & \cellcolor[HTML]{9AFF99}\textbf{100} & 0                                    & 0                                    & 0                                    & 0                                    & 0                                   & 0                                    & 0                                    & 0                                    & 0                                    & 0                                    & 0                                    \\ \hline
\textbf{LC7}                    & 0                                    & 0                                    & 0                                    & 0                                    & 0                                    & 0                                    & 0                                    & \cellcolor[HTML]{9AFF99}\textbf{100} & 0                                    & 0                                    & 0                                    & 0                                   & 0                                    & 0                                    & 0                                    & 0                                    & 0                                    & 0                                    \\ \hline
\textbf{LC8}                    & 0                                    & 0                                    & 0                                    & 0                                    & 0                                    & 0                                    & 0                                    & 0                                    & \cellcolor[HTML]{9AFF99}\textbf{100} & 0                                    & 0                                    & 0                                   & 0                                    & 0                                    & 0                                    & 0                                    & 0                                    & 0                                    \\ \hline
\textbf{LC9}                    & 0                                    & 0                                    & 0                                    & 0                                    & 0                                    & 0                                    & 0                                    & 0                                    & 0                                    & \cellcolor[HTML]{9AFF99}\textbf{100} & 0                                    & 0                                   & 0                                    & 0                                    & 0                                    & 0                                    & 0                                    & 0                                    \\ \hline
\textbf{LC10}                   & 0                                    & 0                                    & 0                                    & 0                                    & 0                                    & 0                                    & 0                                    & 0                                    & 0                                    & 0                                    & \cellcolor[HTML]{9AFF99}\textbf{100} & 0                                   & 0                                    & 0                                    & 0                                    & 0                                    & 0                                    & 0                                    \\ \hline
\textbf{LC11}                   & 0                                    & 0                                    & 0                                    & 0                                    & 0                                    & 0                                    & 0                                    & 0                                    & 0                                    & 0                                    & \cellcolor[HTML]{FE996B}28           & \cellcolor[HTML]{9AFF99}\textbf{72} & 0                                    & 0                                    & 0                                    & 0                                    & 0                                    & 0                                    \\ \hline
\textbf{IE1}                    & 0                                    & 0                                    & 0                                    & 0                                    & 0                                    & 0                                    & 0                                    & 0                                    & 0                                    & 0                                    & 0                                    & 0                                   & \cellcolor[HTML]{9AFF99}\textbf{100} & 0                                    & 0                                    & 0                                    & 0                                    & 0                                    \\ \hline
\textbf{IE2}                    & 0                                    & 0                                    & 0                                    & 0                                    & 0                                    & 0                                    & 0                                    & 0                                    & 0                                    & 0                                    & 0                                    & 0                                   & 0                                    & \cellcolor[HTML]{9AFF99}\textbf{100} & 0                                    & 0                                    & 0                                    & 0                                    \\ \hline
\textbf{LH1}                    & 0                                    & 0                                    & 0                                    & 0                                    & 0                                    & 0                                    & 0                                    & 0                                    & 0                                    & 0                                    & 0                                    & 0                                   & 0                                    & 0                                    & \cellcolor[HTML]{9AFF99}\textbf{100} & 0                                    & 0                                    & 0                                    \\ \hline
\textbf{LH2}                    & 0                                    & 0                                    & 0                                    & 0                                    & 0                                    & 0                                    & 0                                    & 0                                    & 0                                    & 0                                    & 0                                    & 0                                   & 0                                    & 0                                    & 0                                    & \cellcolor[HTML]{9AFF99}\textbf{100} & 0                                    & 0                                    \\ \hline
\textbf{LK1}                    & 0                                    & 0                                    & 0                                    & 0                                    & 0                                    & 0                                    & 0                                    & \cellcolor[HTML]{FE0000}100          & 0                                    & 0                                    & 0                                    & 0                                   & 0                                    & 0                                    & 0                                    & 0                                    & \cellcolor[HTML]{9AFF99}\textbf{0}   & 0                                    \\ \hline
\textbf{LR1}                    & 0                                    & 0                                    & 0                                    & 0                                    & 0                                    & 0                                    & 0                                    & 0                                    & 0                                    & 0                                    & 0                                    & 0                                   & 0                                    & 0                                    & 0                                    & 0                                    & 0                                    & \cellcolor[HTML]{9AFF99}\textbf{100} \\ \hline
\multicolumn{19}{|c|}{\textbf{Times New Roman (T)}}                                                                                                                                                                                                                                                                                                                                                                                                                                                                                                                                                                                                                                                                                                          \\ \hline
\textbf{LB1}                    & \cellcolor[HTML]{9AFF99}\textbf{100} & 0                                    & 0                                    & 0                                    & 0                                    & 0                                    & 0                                    & 0                                    & 0                                    & 0                                    & 0                                    & 0                                   & 0                                    & 0                                    & 0                                    & 0                                    & 0                                    & 0                                    \\ \hline
\textbf{LC1}                    & 0                                    & \cellcolor[HTML]{9AFF99}\textbf{100} & 0                                    & 0                                    & 0                                    & 0                                    & 0                                    & 0                                    & 0                                    & 0                                    & 0                                    & 0                                   & 0                                    & 0                                    & 0                                    & 0                                    & 0                                    & 0                                    \\ \hline
\textbf{LC2}                    & 0                                    & 0                                    & \cellcolor[HTML]{9AFF99}\textbf{100} & 0                                    & 0                                    & 0                                    & 0                                    & 0                                    & 0                                    & 0                                    & 0                                    & 0                                   & 0                                    & 0                                    & 0                                    & 0                                    & 0                                    & 0                                    \\ \hline
\textbf{LC3}                    & 0                                    & 0                                    & 0                                    & \cellcolor[HTML]{9AFF99}\textbf{100} & 0                                    & 0                                    & 0                                    & 0                                    & 0                                    & 0                                    & 0                                    & 0                                   & 0                                    & 0                                    & 0                                    & 0                                    & 0                                    & 0                                    \\ \hline
\textbf{LC4}                    & 0                                    & 0                                    & 0                                    & 0                                    & \cellcolor[HTML]{9AFF99}\textbf{100} & 0                                    & 0                                    & 0                                    & 0                                    & 0                                    & 0                                    & 0                                   & 0                                    & 0                                    & 0                                    & 0                                    & 0                                    & 0                                    \\ \hline
\textbf{LC5}                    & 0                                    & 0                                    & 0                                    & 0                                    & 0                                    & \cellcolor[HTML]{9AFF99}\textbf{100} & 0                                    & 0                                    & 0                                    & 0                                    & 0                                    & 0                                   & 0                                    & 0                                    & 0                                    & 0                                    & 0                                    & 0                                    \\ \hline
\textbf{LC6}                    & 0                                    & 0                                    & 0                                    & 0                                    & 0                                    & 0                                    & \cellcolor[HTML]{9AFF99}\textbf{100} & 0                                    & 0                                    & 0                                    & 0                                    & 0                                   & 0                                    & 0                                    & 0                                    & 0                                    & 0                                    & 0                                    \\ \hline
\textbf{LC7}                    & 0                                    & 0                                    & 0                                    & 0                                    & 0                                    & 0                                    & 0                                    & \cellcolor[HTML]{9AFF99}\textbf{100} & 0                                    & 0                                    & 0                                    & 0                                   & 0                                    & 0                                    & 0                                    & 0                                    & 0                                    & 0                                    \\ \hline
\textbf{LC8}                    & 0                                    & 0                                    & 0                                    & 0                                    & 0                                    & 0                                    & 0                                    & 0                                    & \cellcolor[HTML]{9AFF99}\textbf{100} & 0                                    & 0                                    & 0                                   & 0                                    & 0                                    & 0                                    & 0                                    & 0                                    & 0                                    \\ \hline
\textbf{LC9}                    & 0                                    & 0                                    & 0                                    & 0                                    & 0                                    & 0                                    & 0                                    & 0                                    & 0                                    & \cellcolor[HTML]{9AFF99}\textbf{100} & 0                                    & 0                                   & 0                                    & 0                                    & 0                                    & 0                                    & 0                                    & 0                                    \\ \hline
\textbf{LC10}                   & 0                                    & 0                                    & 0                                    & 0                                    & 0                                    & 0                                    & 0                                    & 0                                    & 0                                    & 0                                    & \cellcolor[HTML]{9AFF99}\textbf{100} & 0                                   & 0                                    & 0                                    & 0                                    & 0                                    & 0                                    & 0                                    \\ \hline
\textbf{LC11}                   & 0                                    & 0                                    & 0                                    & 0                                    & 0                                    & 0                                    & 0                                    & 0                                    & 0                                    & 0                                    & \cellcolor[HTML]{FE0000}100          & \cellcolor[HTML]{9AFF99}\textbf{0}  & 0                                    & 0                                    & 0                                    & 0                                    & 0                                    & 0                                    \\ \hline
\textbf{IE1}                    & 0                                    & 0                                    & 0                                    & 0                                    & 0                                    & 0                                    & 0                                    & 0                                    & 0                                    & 0                                    & 0                                    & 0                                   & \cellcolor[HTML]{9AFF99}\textbf{100} & 0                                    & 0                                    & 0                                    & 0                                    & 0                                    \\ \hline
\textbf{IE2}                    & 0                                    & 0                                    & 0                                    & 0                                    & 0                                    & 0                                    & 0                                    & 0                                    & 0                                    & 0                                    & 0                                    & 0                                   & 0                                    & \cellcolor[HTML]{9AFF99}\textbf{100} & 0                                    & 0                                    & 0                                    & 0                                    \\ \hline
\textbf{LH1}                    & 0                                    & 0                                    & 0                                    & 0                                    & 0                                    & 0                                    & 0                                    & 0                                    & 0                                    & 0                                    & 0                                    & 0                                   & 0                                    & 0                                    & \cellcolor[HTML]{9AFF99}\textbf{100} & 0                                    & 0                                    & 0                                    \\ \hline
\textbf{LH2}                    & 0                                    & 0                                    & 0                                    & 0                                    & 0                                    & 0                                    & 0                                    & 0                                    & 0                                    & 0                                    & 0                                    & 0                                   & 0                                    & 0                                    & \cellcolor[HTML]{FE0000}96           & \cellcolor[HTML]{9AFF99}\textbf{4}   & 0                                    & 0                                    \\ \hline
\textbf{LK1}                    & 0                                    & 0                                    & 0                                    & 0                                    & 0                                    & 0                                    & 0                                    & 0                                    & 0                                    & 0                                    & 0                                    & 0                                   & 0                                    & 0                                    & 0                                    & 0                                    & \cellcolor[HTML]{9AFF99}\textbf{100} & 0                                    \\ \hline
\textbf{LR1}                    & 0                                    & 0                                    & 0                                    & 0                                    & 0                                    & 0                                    & 0                                    & 0                                    & 0                                    & 0                                    & 0                                    & 0                                   & 0                                    & 0                                    & 0                                    & 0                                    & 0                                    & \cellcolor[HTML]{9AFF99}\textbf{100} \\ \hline
\end{tabular}
}
\end{table}

%% file: Tables/DB2_ConfMat_18P_S.tex
\begin{table}[htb!]
\caption{Mean confusion matrices (in \%) for Comic Sans font under cross font scenario using all printers in DB2\cite{joshi2018single} obtained with grid pooling and correlation-based prediction.}
\label{tab:DB2_ConfMat_18P_S}
\resizebox{\textwidth}{!}{
\begin{tabular}{|c|c|c|c|c|c|c|c|c|c|c|c|c|c|c|c|c|c|c|}
\hline
\multicolumn{19}{|c|}{\textbf{Comic Sans (S)}}                                                                                                                                                                                                                                                                                                                                                                                                                                                                                                                                                                                                                                                                                                               \\ \hline
\textbf{LB1}                    & \cellcolor[HTML]{9AFF99}\textbf{100} & 0                                    & 0                                    & 0                                    & 0                                    & 0                                    & 0                                    & 0                                    & 0                                    & 0                                    & 0                                    & 0                                   & 0                                    & 0                                    & 0                                    & 0                                    & 0                                    & 0                                    \\ \hline
\textbf{LC1}                    & 0                                    & \cellcolor[HTML]{9AFF99}\textbf{100} & 0                                    & 0                                    & 0                                    & 0                                    & 0                                    & 0                                    & 0                                    & 0                                    & 0                                    & 0                                   & 0                                    & 0                                    & 0                                    & 0                                    & 0                                    & 0                                    \\ \hline
\textbf{LC2}                    & 0                                    & 0                                    & \cellcolor[HTML]{9AFF99}\textbf{0}   & 0                                    & 0                                    & 0                                    & 0                                    & 0                                    & 0                                    & 0                                    & 0                                    & 0                                   & 0                                    & 0                                    & \cellcolor[HTML]{FE0000}100          & 0                                    & 0                                    & 0                                    \\ \hline
\textbf{LC3}                    & 0                                    & 0                                    & 0                                    & \cellcolor[HTML]{9AFF99}\textbf{0}   & 0                                    & 0                                    & \cellcolor[HTML]{FFFC9E}8            & 0                                    & 0                                    & 0                                    & 0                                    & 0                                   & 0                                    & 0                                    & \cellcolor[HTML]{FFFE65}20           & \cellcolor[HTML]{FE0000}72           & 0                                    & 0                                    \\ \hline
\textbf{LC4}                    & 0                                    & 0                                    & 0                                    & 0                                    & \cellcolor[HTML]{9AFF99}\textbf{0}   & 0                                    & \cellcolor[HTML]{FFFE65}20           & 0                                    & 0                                    & 0                                    & 0                                    & 0                                   & 0                                    & 0                                    & \cellcolor[HTML]{FE0000}80           & 0                                    & 0                                    & 0                                    \\ \hline
\textbf{LC5}                    & 0                                    & 0                                    & 0                                    & 0                                    & 0                                    & \cellcolor[HTML]{9AFF99}\textbf{100} & 0                                    & 0                                    & 0                                    & 0                                    & 0                                    & 0                                   & 0                                    & 0                                    & 0                                    & 0                                    & 0                                    & 0                                    \\ \hline
\textbf{LC6}                    & 0                                    & 0                                    & 0                                    & 0                                    & 0                                    & 0                                    & \cellcolor[HTML]{9AFF99}\textbf{0}   & 0                                    & 0                                    & 0                                    & 0                                    & 0                                   & 0                                    & 0                                    & \cellcolor[HTML]{FE0000}100          & 0                                    & 0                                    & 0                                    \\ \hline
\textbf{LC7}                    & 0                                    & 0                                    & 0                                    & 0                                    & 0                                    & 0                                    & 0                                    & \cellcolor[HTML]{9AFF99}\textbf{100} & 0                                    & 0                                    & 0                                    & 0                                   & 0                                    & 0                                    & 0                                    & 0                                    & 0                                    & 0                                    \\ \hline
\textbf{LC8}                    & 0                                    & 0                                    & 0                                    & 0                                    & 0                                    & 0                                    & 0                                    & 0                                    & \cellcolor[HTML]{9AFF99}\textbf{4}   & 0                                    & 0                                    & 0                                   & 0                                    & 0                                    & \cellcolor[HTML]{FE0000}96           & 0                                    & 0                                    & 0                                    \\ \hline
\textbf{LC9}                    & 0                                    & 0                                    & 0                                    & 0                                    & 0                                    & 0                                    & \cellcolor[HTML]{FE996B}24           & \cellcolor[HTML]{FFFE65}20           & 0                                    & \cellcolor[HTML]{9AFF99}\textbf{56}  & 0                                    & 0                                   & 0                                    & 0                                    & 0                                    & 0                                    & 0                                    & 0                                    \\ \hline
\textbf{LC10}                   & 0                                    & 0                                    & 0                                    & 0                                    & 0                                    & 0                                    & 0                                    & 0                                    & 0                                    & 0                                    & \cellcolor[HTML]{9AFF99}\textbf{100} & 0                                   & 0                                    & 0                                    & 0                                    & 0                                    & 0                                    & 0                                    \\ \hline
\textbf{LC11}                   & 0                                    & 0                                    & 0                                    & 0                                    & 0                                    & 0                                    & 0                                    & 0                                    & 0                                    & 0                                    & \cellcolor[HTML]{FE0000}52           & \cellcolor[HTML]{9AFF99}\textbf{48} & 0                                    & 0                                    & 0                                    & 0                                    & 0                                    & 0                                    \\ \hline
\textbf{IE1}                    & 0                                    & 0                                    & 0                                    & 0                                    & 0                                    & 0                                    & 0                                    & 0                                    & 0                                    & 0                                    & 0                                    & 0                                   & \cellcolor[HTML]{9AFF99}\textbf{100} & 0                                    & 0                                    & 0                                    & 0                                    & 0                                    \\ \hline
\textbf{IE2}                    & 0                                    & 0                                    & 0                                    & 0                                    & 0                                    & 0                                    & 0                                    & 0                                    & 0                                    & 0                                    & 0                                    & 0                                   & \cellcolor[HTML]{FFFC9E}4            & \cellcolor[HTML]{9AFF99}\textbf{96}  & 0                                    & 0                                    & 0                                    & 0                                    \\ \hline
\textbf{LH1}                    & 0                                    & 0                                    & 0                                    & 0                                    & 0                                    & 0                                    & 0                                    & 0                                    & 0                                    & 0                                    & 0                                    & 0                                   & 0                                    & 0                                    & \cellcolor[HTML]{9AFF99}\textbf{100} & 0                                    & 0                                    & 0                                    \\ \hline
\textbf{LH2}                    & 0                                    & 0                                    & 0                                    & 0                                    & 0                                    & 0                                    & 0                                    & 0                                    & 0                                    & 0                                    & 0                                    & 0                                   & 0                                    & 0                                    & \cellcolor[HTML]{FE0000}80           & \cellcolor[HTML]{9AFF99}\textbf{20}  & 0                                    & 0                                    \\ \hline
\textbf{LK1}                    & 0                                    & 0                                    & 0                                    & 0                                    & 0                                    & 0                                    & 0                                    & \cellcolor[HTML]{FE0000}88           & 0                                    & 0                                    & 0                                    & 0                                   & 0                                    & 0                                    & \cellcolor[HTML]{FFFE65}12           & 0                                    & \cellcolor[HTML]{9AFF99}\textbf{0}   & 0                                    \\ \hline
\textbf{LR1}                    & 0                                    & 0                                    & 0                                    & 0                                    & 0                                    & 0                                    & 0                                    & 0                                    & 0                                    & 0                                    & 0                                    & 0                                   & 0                                    & 0                                    & 0                                    & 0                                    & 0                                    & \cellcolor[HTML]{9AFF99}\textbf{100} \\ \hline
\end{tabular}
}
\end{table}

%% file: Sections/S5_Conclusion.tex
\section{Conclusion}
\label{sec:Conclusion}
We have proposed a new printer-specific pooling based system for source printer identification.
We use an extensive set of visual analysis and experimental results to show that a printer's signature varies across a printed document, and this variation can be captured by dividing the document image into non-overlapping blocks.
We analyzed two variants for block pooling, i.e., column and grid pooling.
For document images scanned at lower precision (i.e., 8-bit depth), we use SVM-based prediction (as used in most state-of-the-art methods).
The proposed method performs better than most methods on the publicly available dataset and performs similar to the PSLTD-based method.
Results show that column pooling performs better than grid pooling using an ensemble of SVMs as the number of samples available for each SVM (corresponding to each block) is insufficient to learn a discriminative model.
However, for document images scanned at higher precision (i.e., 16-bit depth), the proposed method's discriminative power is highlighted by a correlation-based prediction instead of a complex classifier.
Both grid and column variants of the proposed method performs better than state-of-the-art methods when evaluated under cross font scenario.
An extensive set of experiments reveal that the grid pooling variant performs better than column pooling using a correlation-based technique.
The proposed method's grid pooling variant achieves more than 93.5\% and 94.3\%accuracies when tested on documents containing  Arial and Times New Roman font types and trained using documents containing only Cambria font type.
Classification with Pearson correlation provides promising results that could pave the way for classifier independent detection approach with a high precision image acquisition process.

The proposed method and the existing state-of-the-art combination of PSLTD and consecutive pooling does not perform well on Comic Sans font as it is drastically different from the other three font types in our dataset DB2.
The proposed method requires that the test document contains a sufficient number of letters printed in a block to estimate a good quality printer signature.
The proposed method is less useful for intra-page forgery locations, as its localization ability is reduced if the prediction of source printer is carried out on fewer pooled feature vectors.
There are other open challenges that need to be analyzed in detail, including toner variation, paper quality and type, and printer age.
Future work will include improving the printer signature model to address cross-font type, cross-font size, and cross-language scenarios. It would allow the scaling of source printer identification systems to a larger number of printers.